\documentclass[prd,onecolumn,nofootinbib,aps,tightenlines]{revtex4-1}

\usepackage{amsmath}
\usepackage{amsfonts}
\usepackage{amssymb}
\usepackage{dsfont}
\usepackage{mathtools}
\usepackage{tabularx}
\usepackage{enumerate}
\usepackage{array}
\usepackage{color}
\usepackage{graphicx}
\usepackage{floatrow}
\usepackage{hyperref}

\newcommand{\units}[1]{\ensuremath{\mathrm{#1}}}
\newcommand{\GeV}{\units{GeV}}
\newcommand{\MeV}{\units{MeV}}
\newcommand{\keV}{\units{keV}}
\newcommand{\km}{\units{km}}
\newcommand{\cm}{\units{cm}}
\newcommand{\s}{\units{s}}
\newcommand{\Gyr}{\units{Gyr}}
\newcommand{\g}{\units{g}}

\newcommand{\msolar}{\ensuremath{M_\odot}}
\newcommand{\ket}[1]{\left|#1\right\rangle}
\newcommand{\mH}{m_\textrm{H}}
\newcommand{\sigVisc}{\sigma_V}
\newcommand{\siginel}{\sigma_\textrm{++}}
\newcommand{\Ehf}{E_\textrm{hf}}
\newcommand{\vrms}{v_\textrm{rms}}

\newcolumntype{L}[1]{>{\hsize=#1\hsize\raggedright\arraybackslash}X}%
\newcolumntype{R}[1]{>{\hsize=#1\hsize\raggedleft\arraybackslash}X}%
\newcolumntype{C}[1]{>{\hsize=#1\hsize\centering\arraybackslash}X}%

\newfloatcommand{capbtabbox}{table}[][\FBwidth]

\graphicspath{ {../Figures/}, {figures/} }

\begin{document}

\title{Hidden sector hydrogen as dark matter: Small-scale structure formation predictions and the importance of hyperfine interactions}
\author{Kimberly K.~Boddy}
\affiliation{Department of Physics and Astronomy, University of Hawaii,
  Honolulu, Hawaii 96822, USA}
\author{Manoj Kaplinghat}
\author{Anna Kwa}
\affiliation{Department of Physics and Astronomy, University of California,
  Irvine, California 92697, USA}
\author{Annika H.~G.~Peter}
\affiliation{Department of Physics and Astronomy,
  Center for Cosmology and AstroParticle Physics, Ohio State University,
  Columbus, Ohio 43210, USA}

\begin{abstract}
We study the atomic physics and the astrophysical implications of a model in which the dark matter is the analog of hydrogen in a secluded sector.
The self-interactions between dark matter particles include both elastic scatterings as well as inelastic processes due to a hyperfine transition.
The self-interaction cross sections are computed by numerically solving the coupled Schr\"{o}dinger equations for this system.
We show that these self-interactions exhibit the right velocity dependence to explain the low dark matter density cores seen in small galaxies while being consistent with all constraints from observations of clusters of galaxies.
For a viable solution, the dark hydrogen mass has to be in 10--100 GeV range and the dark fine-structure constant has to be larger than 0.01.
This range of model parameters requires the existence of a dark matter--anti-matter asymmetry in the early universe to set the relic abundance of dark matter.
For this range of parameters, we show that significant cooling losses may occur due to inelastic excitations to the hyperfine state and subsequent decays, with implications for the evolution of low-mass halos and the early growth of supermassive black holes.
Cooling from excitations to higher $n$ levels of dark hydrogen and subsequent decays is possible at the cluster scale, with a strong dependence on halo mass.
Finally, we show that the minimum halo mass is in the range of $10^{3.5}$ to $10^7 M_\odot$ for the viable regions of parameter space, significantly larger than the typical predictions for weakly interacting dark matter models.
This pattern of observables in cosmological structure formation is unique to this model, making it possible to rule in or rule out hidden sector hydrogen as a viable dark matter model.
\end{abstract}
\maketitle

\section{Introduction}

While the standard $\Lambda$CDM cosmological model with collisionless, cold dark matter (CDM) is successful in explaining the observed large-scale structure in the Universe, there are many puzzles on galactic scales yet to be explained convincingly by a CDM-based scenario.
$N$-body simulations of CDM structure growth predict cuspy radial density profiles ($\rho\sim r^{-1}$) with high central densities~\cite{Navarro:1995iw,Bullock:1999he,Diemand:2007qr,Stadel:2008pn}; however, observations of rotation curves reveal cored ($\rho\sim$ constant) or otherwise low-density inner regions in dark matter dominated galaxies, from dwarfs to low surface brightness (LSB) galaxies ~\cite{Moore:1994yx,Persic:1995ru,deBlok:1996jib,deBlok:2001hbg,Swaters:2002rx,Simon:2004sr,Spekkens:2005ik,KuziodeNaray:2007qi,deBlok:2008wp,Donato:2009ab,Oh:2010ea,Adams:2014bda}.
Galaxy clusters also show evidence for a deficit of dark matter within the effective stellar radius of the central galaxy, with the mass profile outside being consistent with CDM predictions~\cite{Newman:2009qm, Newman:2011ip, Newman:2012nw}.
The dark matter density profiles in dwarf spheroidal galaxies (dSphs) are a subject of current debate, with various studies finding that the stellar data for various dSphs is most consistent with a core~\cite{Walker:2011zu,Amorisco:2011hb,Jardel:2011yh,Amorisco:2012rd,Amorisco:2013uwa}, or a cusp~\cite{Jardel:2012am,Richardson:2014mra}, or is unconstrained~\cite{Breddels:2013qqh}.
However, it seems clear that these galaxies are less dense than expected in pure CDM models~\cite{BoylanKolchin:2011de,BoylanKolchin:2011dk,Papastergis:2014aba}.
The inclusion of supernovae and/or black hole feedback processes in cosmological simulations may ameliorate these anomalies in dwarf galaxies by significantly altering the central gravitational potential~\cite{Governato:2012fa,Brooks:2012vi,Maccio:2006nu,Stadel:2008pn,Oh:2010mc,Onorbe:2015ija,Katz:2016hyb,Read:2015sta,Wetzel:2016wro}.
However, it is unclear whether such effects are simultaneously able to affect the halo structure to the extent observed in low-mass ($M_\ast \sim10^6$--$10^7 \msolar$) isolated dwarf galaxies~\cite{Ferrero:2011au,Weinberg:2013aya,Papastergis:2015tg} and low surface brightness galaxies~\cite{deNaray:2011hy}.
It is also not clear if the diversity of cores observed inferred from rotation curves can be explained in the context of these models \cite{deNaray:2009xj,Oman:2015xda,Pace:2016oim}.

Self-interacting dark matter (SIDM) is an attractive solution~\cite{Spergel:1999mh} to these anomalies that works across the range of mass scales under consideration, from dwarf galaxies to galaxy clusters~\cite{Rocha:2012jg}.
In such a scenario, scatterings between dark matter particles allow for energy to be transferred from the hotter outer regions of the halo into the colder innermost regions.
SIDM halos thus have hotter cores with higher velocity dispersions than the cold, cuspy interiors of collisionless dark matter halos, which lack a mechanism to heat the inner cusp into a core.
$N$-body simulations~\cite{Vogelsberger:2012ku,Rocha:2012jg} and analytic models based on these simulations find that the aforementioned issues in small-scale structure (on the dSph and LSB scales) may be alleviated if the dark matter is strongly self-interacting with a hard-sphere scattering cross section of $0.5~\cm^2/\g \lesssim \sigma /m \lesssim 5.0~\cm^2/\g$~\cite{Zavala:2012us,Elbert:2014bma,Kaplinghat:2015aga,Vogelsberger:2015gpr}, where $m$ is the mass of the dark matter.
In order to produce cores of radius $10$--$50$ kpc in cluster-sized halos in which the relative dark matter particle velocity is $v\sim 1000~\km/\s$, the required cross section is $\sigma\sim 0.1~\cm^2/\g$~\cite{Kaplinghat:2015aga}.
We are thus motivated to consider SIDM models with \textit{velocity-dependent} cross sections that are suppressed at cluster-scale velocities.
Upper limits on the SIDM cross section may also be derived through the observed ellipticities of cluster-scale halos ($\sigma/m \lesssim 1~\cm^2/\g$)~\cite{Peter:2012jh} and the measured center-of-mass offsets in merging cluster systems ($\sigma/m<0.47~\cm^2/\g$)~\cite{Harvey:2015hha}, but we find both these constraints to be weaker than those obtained from measurements of the inner density profiles of galaxy clusters~\cite{Kaplinghat:2015aga}.

Atomic dark matter~\cite{Goldberg:1986nk, Mohapatra:2001sx, Feng:2009mn, Kaplan:2009de, Behbahani:2010xa, CyrRacine:2012fz, Cline:2013pca,Foot:2014uba}, in which the features of Standard Model (SM) hydrogen are copied to a dark sector, has all the features required of an SIDM model to solve the small-scale puzzles.
We consider a dark proton and dark electron, which are charged under an unbroken U(1) gauge group and may combine to form dark hydrogen.
If the formation of dark hydrogen bound states is efficient, these dark atoms constitute approximately all of the dark matter.
Since the dark hydrogen is a composite particle with an extended, finite size, its self-interaction cross section can be naturally large, as required for SIDM.

In this work, we consider atomic dark matter that exists today exclusively in bound states---dark recombination was fully complete.
This model has uniquely testable phenomenology due to the ability of atomic dark matter to dissipate energy.
We calculate and explore the cosmological consequences of both collisional scattering (which transfers energy between between dark atoms) and inelastic hyperfine upscattering (which results in energy loss through excitations and subsequent decays).
These cross sections are velocity dependent, allowing the self-interactions to modify the halo profile to varying degrees in different astrophysical systems.
The general trend is that the cross sections of particles in dwarf halos with characteristic velocities of $\sim 40~\km/\s$ will be larger than those in cluster halos with characteristic velocities of $\sim 1000~\km/\s$, which allows for regions of parameter space in which this model may resolve the aforementioned issues in small-scale structure.
The heating rate from scatterings as well as the cooling losses from inelastic collisions vary widely depending on both the model parameters and the radial position in a halo of a given mass.
The combined effects of both types of scatterings may thus lead to nontrivial effects on dark matter halo structure and evolution.
At higher particle energies, additional atomic interactions, such as collisional excitations to the $n=2$ state and ionization, may begin to affect the structure of cluster-sized halos.

For this interesting range of parameter space, where we see competing effects from collisional heating and cooling processes on the evolution of halos, we find additional features in the small-scale halo mass function that allow us to distinguish atomic dark matter from CDM cosmologically.
Coupling between the dark matter and dark radiation produces dark acoustic oscillations, which are weakly constrained by measurements of the matter power spectrum and the cosmic microwave background (CMB)~\cite{CyrRacine:2013fsa}.
The most interesting effect of acoustic oscillations in the dark plasma would be a cutoff in the matter power spectrum set by the size of the fluctuations entering the horizon before the time of matter-radiation decoupling, resulting in a minimum dark matter halo mass that is significantly larger than in the typical weak-scale models without hidden sectors.

Altering small-scale structure with SIDM neither assumes nor requires any interactions between dark matter and SM particles beyond gravitational interactions; thus, we take a minimal approach and seclude the dark sector from the visible sector.
Atomic dark matter has been presented in other contexts, such as a mirror universe~\cite{Mohapatra:2001sx,Foot:2014mia} and supersymmetry~\cite{Behbahani:2010xa}.
The effect of hidden sector dissipative dark matter on small-scale structure has been previously studied in Refs.~\cite{Foot:2014uba, Foot:2014mia}.
We note that our approach differs from prior works~\cite{Foot:2014uba, Foot:2014mia}: the cored profiles in this work result from the collisional scatterings of the neutral dark atoms, whereas the density profiles in Ref.~\cite{Foot:2014uba} are shaped by a combination of bremsstrahlung cooling processes in the dark sector as well as energy injection from visible supernovae [which is made possible through the inclusion of a kinetic mixing interaction between the dark U(1) and the SM hypercharge].
Kinetic mixing has also been used to explain DAMA and CoGeNT~\cite{Kaplan:2009de,Kaplan:2011yj,Cline:2012ei,Cline:2012is,Foot:2014uba} and the $3.5~\keV$ line~\cite{Cline:2014eaa}.

The paper is structured as follows.
In Sec.~\ref{sec:model} we describe the atomic dark matter model and the scattering properties of dark hydrogen.
In Sec.~\ref{sec:apps} we consider dark hydrogen as an SIDM candidate and determine the parameter space allowed to accommodate SIDM and satisfy cosmological constraints.
In Sec.~\ref{sec:inelastic} we investigate how inelastic scattering processes can affect halo formation.
In Sec.~\ref{sec:clusters} we discuss the possibility of upscatterings to the $n=2$ excited state as well as collisional ionization in cluster-scale halos.
We conclude in Sec.~\ref{sec:conclusions}.

\section{Atomic Dark Matter Model}
\label{sec:model}

We begin this section by describing the properties and parameters of a secluded dark atomic model.
We then present the formalism for dark hydrogen-hydrogen scattering and show results for the scattering cross section, obtained by numerically solving the Schr\"{o}dinger equation.
Our calculations agree with a previous detailed study for elastic scattering~\cite{Cline:2013pca}.
We further improve upon the basic model by incorporating inelastic processes that arise from hyperfine interactions.
In subsequent sections, we show that the hyperfine splitting of ground-state dark hydrogen provides a rich phenomenological framework to study structure formation.

We consider dark matter in a secluded sector that mimics the properties of SM hydrogen.
The dark sector has two elementary particles: a dark electron with mass $m_e$ and a dark proton with mass $m_p$.
These particles have opposite charge under an unbroken U(1) gauge, and they interact with a strength given by a dark fine-structure constant $\alpha$.
The dark electron and dark proton may combine to form a neutral dark hydrogen atom with mass $m_H$ and a binding energy $B_\mathrm{H} = \alpha^2 \mu_H/2$, where $\mu_H$ is the reduced mass of the dark electron-proton system.
It is convenient to parametrize the theory in terms of the following:
\begin{equation}
  \mu_H = \frac{m_e m_p}{m_e+m_p} \ , \qquad
  R \equiv \frac{m_p}{m_e} \ , \qquad
  f(R,\alpha) \equiv \frac{m_H}{\mu_H} = R+2+\frac{1}{R}-\frac{\alpha^2}{2} \ .
  \label{eq:parameters}
\end{equation}
Without loss of generality, we set $R>1$.
If $R \gg 1$, we may ignore the contribution from the binding energy so that $f(R)$ is a function of $R$ only,
\begin{equation}
  f(R) \approx R + 2 + R^{-1} \ .
\end{equation}

All mass and coupling variables refer to dark-sector quantities and are not used in this paper to refer to their visible-sector counterparts.
However, since the analysis in this section closely follows that of SM hydrogen, dark hydrogen can have the same atomic properties of SM hydrogen by setting the model parameters appropriately.
In fact, we can further diminish the distinction between the generic dark atomic model and SM hydrogen by expressing all dimensionful quantities in terms of the atomic energy and length scales
\begin{equation}
  E_0 = \alpha^2 \mu_H \ , \qquad
  a_0 = (\alpha \mu_H)^{-1} \ .
  \label{eq:atomic-scales}
\end{equation}
In doing so, we may adapt numerical results from the visible sector to the dark sector by adjusting these atomic scales.
For the remainder of this section, we express quantities in terms of atomic units for full generality.

\subsection{Quantum formalism}

We are interested in the interaction between two $n=1$ ground-state dark hydrogen atoms, where $n$ is the principal quantum number.
The formalism is adapted from SM hydrogen~\cite{Kolos:1964lr,Zygelman:2003lr}, and we outline the procedure here.
An essential tool for simplifying the computation of a molecular wave function is the Born-Oppenheimer (BO) approximation, in which we write the total wave function as
\begin{equation}
  \Psi = \sum_\gamma \psi_\gamma (\mathbf{x}) \,
  \phi_\gamma(\mathbf{x},\mathbf{y}) \ ,
  \label{eq:total-wf}
\end{equation}
where $\mathbf{x}$ is the relative separation (in atomic units) of the dark protons and $\mathbf{y}$ is a collective coordinate (in atomic units) for the dark electrons---we discard the center-of-mass motion.
The subscript $\gamma$ is a shorthand notation for the quantum numbers that define a set of basis states.
Although, in principle, $\gamma$ runs over all possible states, we truncate it to include only $n=1$ atomic ground states.
In writing Eq.~\eqref{eq:total-wf}, we treat $\psi_\gamma(\mathbf{x})$ as the nuclear wave function for the dark proton, and it has no explicit dependence on the motion of the dark electrons.
On the other hand, the electronic wave function $\phi_\gamma(\mathbf{x},\mathbf{y})$ is dependent on the relative position of the dark protons.
Accordingly, the Hamiltonian of the system separates into a part that describes only the relative motion of the protons and a part that encompasses the motion of the electrons and all Coulomb interactions.
To solve the Schr\"{o}dinger equation for $\Psi$, the dark protons are initially held fixed, leaving just the electronic part of the Schr\"{o}dinger equation.
By repeatedly solving for $\phi_\gamma$ under various nuclear configurations, the electronic BO eigenvalues $\epsilon_\gamma(x)$ form a potential energy surface, which depends on the distance between the dark protons.
These eigenvalues receive higher-order corrections from vibrational and rotational nuclear motion and from relativistic electronic motion.
The validity of the BO approximation relies on the dark proton being sufficiently heavier than the dark electron; the error of the approximation is $\sim R^{-3/2}$~\cite{Takahashi:2006lr} and so should be near and below the percent level for $R\gtrsim 20$.

We use existing calculations~\cite{Kolos:1974lr,Silvera:1980zz,Wolniewicz:1993lr} of the BO eigenvalues for SM, ground-state molecular hydrogen.
The states are labeled by the total electronic spin $S$.
The $S=0$ spin-singlet state $X^1 \Sigma_g^+$ has a BO eigenvalue $\epsilon_0(x)$ and positive electronic parity; the $S=1$ spin-triplet state $b^3 \Sigma_u^+$ has a BO eigenvalue $\epsilon_1(x)$ and negative electronic parity.
The explicit forms of $\epsilon_0(x)$ and $\epsilon_1(x)$ that we use are found in Ref.~\cite{Cline:2013pca}, with the exception of $\epsilon_0(x)$ in the range $0.3 < x < 12$, for which we interpolate tabulated results in Ref.~\cite{Wolniewicz:1993lr}.
We identify these states with our electronic states $\phi_\gamma(\mathbf{x},\mathbf{y})$, modulo rotated configurations of the dark protons~\cite{Zygelman:1994lr}.

With the BO eigenvalues at hand, we reincorporate the kinetic energy of the dark protons.
Solving the full Schr\"{o}dinger equation reduces to solving a one-dimensional (1D) Schr\"{o}dinger equation for the relative nuclear motion in the potential $\epsilon_\gamma(x)$.
In ket notation, we write the basis for the total wave function as $\ket{SM_S I M_I}$, where $S$ and $I$ are the total electronic and nuclear spins, respectively, and $M_S$ and $M_I$ are their associated $z$-axis projections.
The label $\gamma$ runs over 16 states (or scattering channels) for two ground-state dark hydrogen atoms.
Since the potential depends only on the nuclear separation $x$, we may expand the nuclear wave function in terms of partial waves,
\begin{equation}
  \psi_\gamma (\mathbf{x}) = \sum_{l,m} x^{-1}
  \left[F_\gamma(x)\right]_l \, Y_{lm}(\theta, \phi) \ ,
  \label{eq:nucl-wf}
\end{equation}
where $[F_\gamma(x)]_l$ is the partial wave radial amplitude.
The Schr\"{o}dinger equation we must solve is
\begin{equation}
  \left\{\frac{d^2}{dx^2} - \frac{l(l+1)}{x^2} + f(R)
  \left[ \mathcal{E} - \epsilon_\gamma(x) \right] \right\} [F_\gamma(x)]_l = 0
\end{equation}
for each channel $\gamma$ with angular momentum $l$ and energy $\mathcal{E}$ (in atomic units).
Employing a more succinct notation, we discard the label $\gamma$ and express the Schr\"{o}dinger equation in vector/matrix notation.
We group the amplitudes into a single vector $\mathbf{F}_l$ of length 16, whose row entries correspond to the various channels $\gamma \leftrightarrow \ket{S M_S I M_I}$.
The potential then becomes a diagonal $16 \times 16$ matrix $\mathbb{V}(x)$, whose entries are $\epsilon_{0,1}(x)$ for corresponding channels with $S=0,1$.
The Schr\"{o}dinger equation becomes
\begin{equation}
  \left\{\frac{d^2}{dx^2} - \frac{l(l+1)}{x^2} +
    f(R) \left[ \mathcal{E} - \mathbb{V}(x) - \mathbb{W} \right] \right\}
  \mathbf{F}_l = 0 \ ,
  \label{eq:SE}
\end{equation}
where we have included an additional constant $16 \times 16$ potential matrix $\mathbb{W}$ in anticipation of the next section, but here we set $\mathbb{W}=0$.

\subsection{Hyperfine interaction}

We now incorporate a hyperfine interaction of the form
\begin{equation}
  \hat{H}_\textrm{hf} = \Ehf
  \left(\hat{\mathbf{I}}_A \cdot \hat{\mathbf{S}}_A +
  \hat{\mathbf{I}}_B \cdot \hat{\mathbf{S}}_B \right) \ ,
\end{equation}
between the nuclear and electronic spins of atoms $A$ and $B$ involved in the scattering process.
The interaction creates an energy splitting
\begin{equation}
  \frac{\Ehf}{E_0} = \frac{2}{3} g_e g_p \alpha^2 \frac{1}{f(R)}
  \label{eq:splitting}
\end{equation}
of the $n=1$ ground state into a hyperfine ground state and a hyperfine excited state.
We set the Land\'{e}-$g$ factors of the dark electron and dark proton to be $g_e=2$ and $g_p=2$, respectively.
The hyperfine excited state is unstable with a decay width
\begin{equation}
 \Gamma=\frac{1}{3}~\frac{\alpha \Ehf^3}{m_e^2} \ .
 \label{eq:decay-width}
\end{equation}

\begin{figure}[t]
  \begin{floatrow}
    \centering
    \ffigbox{%
      \includegraphics[scale=0.5,clip=true,trim=12cm 8cm 10cm 7cm]{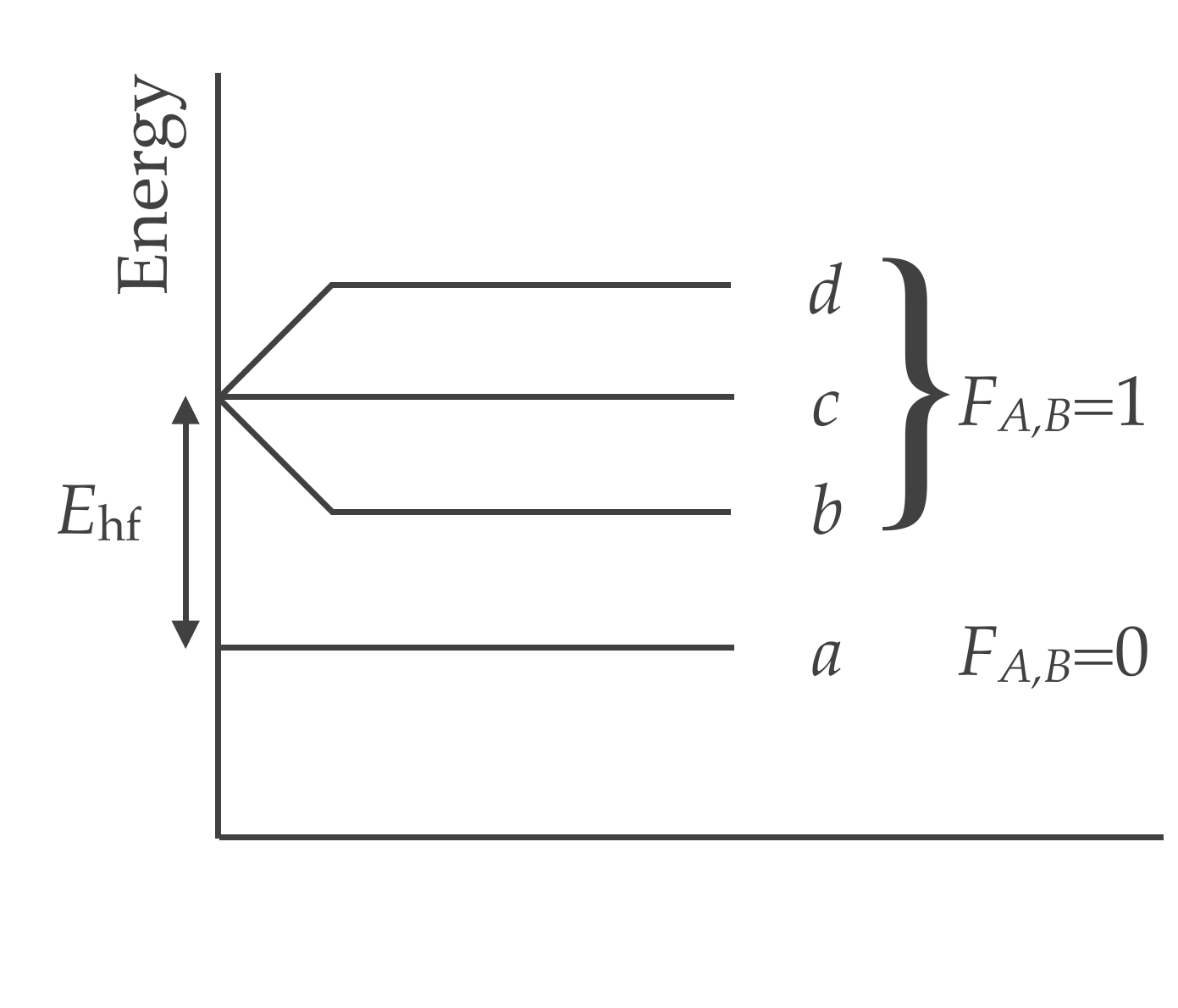}
    }
    {\caption{Energy-level diagram for the $n=1$ ground state of dark hydrogen.
        Hyperfine interactions break the degeneracy between the $F_{A,B}=0$ and $F_{A,B}=1$ states for the dark atoms $A$ and $B$.
        The labels $b$, $c$, and $d$ correspond to $m_F=-1$, $0$, and $1$, respectively.}
      \label{fig:hyperfine-levels}}

    \capbtabbox{%
      \begin{tabularx}{0.5\textwidth}
        {L{2} C{0.5}@{}C{0.5}@{}C{0.5}@{}C{0.5} C{2}}
        Channel & \multicolumn{4}{C{2}}{$\ket{F_A M_A F_B M_B}$} & Level \\
        \hline
        \hline
        1   &  0 & 0 & 0 & 0  &  $aa$ \\
        2   &  1 & 0 & 1 & 0  &  $cc$ \\
        3   &  0 & 0 & 1 & 0  &  $ac$ \\
        4   &  1 & 0 & 0 & 0  &  $ca$ \\
        5   &  1 &-1 & 1 & 1  &  $bd$ \\
        6   &  1 & 1 & 1 &-1  &  $db$ \\
        \hline
        7   &  0 & 0 & 1 & 1  &  $ad$ \\
        8   &  1 & 1 & 0 & 0  &  $da$ \\
        9   &  1 & 0 & 1 & 1  &  $cd$ \\
        10  &  1 & 1 & 1 & 0  &  $dc$ \\
        \hline
        11  &  0 & 0 & 1 &-1  &  $ab$ \\
        12  &  1 &-1 & 0 & 0  &  $ba$ \\
        13  &  1 &-1 & 1 & 0  &  $bc$ \\
        14  &  1 & 0 & 1 &-1  &  $cb$ \\
        \hline
        15  &  1 & 1 & 1 & 1  &  $dd$ \\
        16  &  1 &-1 & 1 &-1  &  $bb$
      \end{tabularx}
    }
    {\caption{List of interaction channels with associated quantum numbers.
        The level for dark atoms $A$ and $B$ corresponds to the labeling in Fig.~\ref{fig:hyperfine-levels}.
        The horizontal lines exhibit the block diagonal nature of the potential $\mathbb{V}(x)$.}
      \label{tab:channels}}
  \end{floatrow}
\end{figure}

The basis $\ket{SM_S I M_I}$ is not ideal for this interaction, so we perform a change of basis~\cite{Zygelman:2003lr} to $\ket{F_AM_AF_BM_B}$, where $\hat{\mathbf{F}}_A=\hat{\mathbf{I}}_A+\hat{\mathbf{S}}_A$ is the total angular momentum of atom $A$ with spin projection $M_A$ (and similarly with atom $B$).
Table \ref{tab:channels} lists the quantum numbers for each channel, and Fig.~\ref{fig:hyperfine-levels} shows a schematic energy-level diagram for the ground state dark hydrogen atom with a hyperfine splitting.
The hyperfine potential
\begin{equation}
  [\mathbb{W}]_{F_A M_A F_B M_B}^{F'_A M'_A F'_B M'_B}
  = \delta_{F'_A F_A} \delta_{F'_B F_B} \delta_{M'_A M_A} \delta_{M'_B M_B}
  \frac{\Ehf}{2E_0} [F_A (F_A+1) + F_B (F_B+1)-3]
\end{equation}
is a diagonal matrix in this basis, while the change of basis induces off-diagonal elements in $\mathbb{V}(x)$.
As a result, the Schr\"{o}dinger equation \eqref{eq:SE} becomes a system of 16 coupled differential equations.
However, the selection rule $\Delta M=0$, where $M=M_A+M_B$, allows $\mathbb{V}(x)$ to be written as a block-diagonal matrix with four%
\footnote{Channels 15 and 16 have different values of $M$, but they are grouped in a single submatrix to maintain consistency with Ref.~\cite{Zygelman:2003lr}.}
submatrices, whose form is given explicitly in Ref.~\cite{Zygelman:2003lr}.
The horizontal lines in Table \ref{tab:channels} indicate which sets of channels correspond to different submatrices.
Thus, we may solve Eq.~\eqref{eq:SE} by solving multiple systems of fewer coupled equations, which is more computationally efficient.

\subsection{Scattering}
\label{subsec:scattering}

The numerical details of solving the Schr\"{o}dinger equation \eqref{eq:SE} are given in Appendix~\ref{sec:numerical}, and we summarize the results here.
The cross section (in atomic units) from the state $\ket{j}\equiv\ket{F_A M_A F_B M_B}$ to $\ket{i}\equiv\ket{F'_A M'_A F'_B M'_B}$ is~\cite{Zygelman:2005lr}:
\begin{equation}
  \sigma(j\to i) = \frac{\pi}{2k_j^2} \sum_l (2l+1)
  \left| (\mathbb{T}_l)_{ij} + (-1)^l(\mathbb{T}_l)_{\tilde{i}j} \right|^2 \ ,
  \label{eq:cross-section}
\end{equation}
where $k_j$ is the wave number (in atomic units), $\mathbb{S}_l = \mathds{1}+\mathbb{T}_l$ is the $S$ matrix, and $\tilde{i}$ denotes swapping the labels on (or the quantum numbers of) dark atoms $A$ and $B$.
The wave number depends on which hyperfine energy level the dark atoms occupy; at a given energy $\mathcal{E}$, a dark atom in the excited state will have a smaller velocity than one in the ground state.
The wave number $k_j$ relates the relative velocity $v_j$ to the asymptotic energy of the channel $j$ via
\begin{equation}
  k_j = \frac{1}{2} f(R) \frac{v_j}{\alpha}
  = \sqrt{f(R)\left[\mathcal{E}-(\mathbb{W})_{jj}\right]} \ .
  \label{eq:wavenumber}
\end{equation}
If there is sufficient energy such that $k_j^2 > 0$, then channel $j$ is open and accessible; otherwise, it is closed.

Instead of using the cross section for specific channels, a more useful quantity is the spin-averaged cross section for a particular type of process.
Based on the change in the total angular momentum $\hat{\mathbf{F}}=\hat{\mathbf{F}}_A+\hat{\mathbf{F}}_B$, we group spin-averaged cross sections into seven categories: $\Delta F=\pm 2$, $\Delta F=\pm 1$, and three separate groups of $\Delta F=0$.
Table \ref{tab:spin-avg} lists these cross sections and their corresponding scattering channels.
Note that the channels in each group have the same initial and final velocities.
The spin-averaged cross sections $\sigma_{+}$ and $\sigma_{++}$ represent one and two dark atoms, respectively, in the hyperfine ground state upscattering to the excited state.
Similarly, $\sigma_{-}$ and $\sigma_{--}$ represent one and two dark atoms downscattering from the excited state to the ground state.
The remaining spin-averaged cross sections involve dark atoms that either do not make an energy transition or simply swap energy states.
Atoms remain in the ground state for $\sigma_{gg}$ and in the excited state for $\sigma_{ee}$.
For $\sigma_{ge}$, one dark atom is in the ground state while the other is in the excited state.
There are some processes not listed in Table \ref{tab:spin-avg}; they are zero due to selection rules or because total angular momentum is not conserved.

\begin{table}[t]
  \centering
  \begin{tabularx}{0.7\textwidth}{L{1}C{1}R{1}@{$\to$}L{1}}
    Cross section & Transition ($\Delta F$)
    & \multicolumn{2}{C{2}}{Processes} \\
    \hline
    \hline
    $\sigma_{++}$ & $2$  & $aa$ & $cc,\ bd,\ db$ \\
    \hline
    $\sigma_{--}$ & $-2$ & $cc,\ bd,\ db$ & $aa$ \\
    \hline
    $\sigma_{+}$  & $1$  & $ac,\ ca$ & $bd,\ db$ \\
    & & $ad,\ da$ & $cd,\ dc$ \\
    & & $ab,\ ba$ & $bc,\ cb$ \\
    \hline
    $\sigma_{-}$  & $-1$ & $bd,\ db$ & $ac,\ ca$ \\
    & & $cd,\ dc$ & $ad,\ da$ \\
    & & $bc,\ cb$ & $ab,\ ba$ \\
    \hline
    $\sigma_{gg}$  & $0$ & $aa$ & $aa$ \\
    \hline
    $\sigma_{ee}$  & $0$ & $cc,\ bd,\ db$ & $cc,\ bd,\ db$ \\
    & & $cd,\ dc$ & $cd,\ dc$ \\
    & & $bc,\ cb$ & $bc,\ cb$ \\
    & & $dd$ & $dd$ \\
    & & $bb$ & $bb$ \\
    \hline
    $\sigma_{ge}$ & $0$ & $ac,\ ca$ & $ac,\ ca$ \\
    & & $ad,\ da$ & $ad,\ da$ \\
    & & $ab,\ ba$ & $ab,\ ba$ \\
    \hline
  \end{tabularx}
  \caption{Definition of various spin-averaged cross sections.}
  \label{tab:spin-avg}
\end{table}

The total spin-averaged cross section $\sigma_\textrm{tot}$ is obtained by summing over all individual cross sections in Eq.~\eqref{eq:cross-section} and dividing by 16, if all channels are open.
There are also variations of the total cross section that incorporate nonuniform weights for the angular integration.
The momentum-transfer cross section $\sigma_{\textrm{tot},T}$ weights the cross section by the fractional longitudinal momentum transferred in the scattering process, thereby suppressing forward scattering; whereas the viscosity cross section $\sigma_{\textrm{tot},V}$ weights the cross section by the fractional transverse energy transfer, thereby suppressing both forward and backward scattering,
\begin{align}
  \sigma_{\textrm{tot},T} &=
  \int\frac{d\sigma_\textrm{tot}}{d\Omega}(1-\cos\theta) d\Omega \\
  \sigma_{\textrm{tot},V} &=
  \int\frac{d\sigma_\textrm{tot}}{d\Omega}\sin^2\theta d\Omega \ .
\end{align}
We may apply these variations to the individual cross sections in Eq.~\eqref{eq:cross-section}, under the assumption that the scattering particles are identical.
For ease of notation, we define $(\mathbb{T}_l^\textrm{eff})_{ij} \equiv (\mathbb{T}_l)_{ij} + (-1)^l(\mathbb{T}_l)_{\tilde{i}j}$.
To obtain $\sigma_V(j\to i)$, we replace $(2l+1) \to (l+1)(l+2)/(2l+3)$ and $\mathbb{T}_l^\textrm{eff} \to \mathbb{T}_{l+2}^\textrm{eff} - \mathbb{T}_{l}^\textrm{eff}$.
The normalization ensures that $\sigma_{\textrm{tot},V}$ has the proper limits for pure $s$-wave scattering, for which the differential cross section is isotropic.
For $\sigma_{\textrm{tot},T}$, however, the part of the integrand with the additional $\cos\theta$ yields zero, so $\sigma_{\textrm{tot},T} = \sigma_\textrm{tot}$~\cite{Krstic:1999lr}.
The suppression of forward scattering is compensated by the identical backward scattering, resulting in no change from $\sigma_\textrm{tot}$.

\section{Applications of Atomic Dark Matter}
\label{sec:apps}

We now examine the atomic dark matter model of Sec.~\ref{sec:model} in a cosmological and astrophysical context.
First, we determine the necessary conditions for dark matter to be in the form of $n=1$ ground state dark hydrogen.
Then, with calculations of scattering cross sections from the previous section at hand, we examine the atomic dark matter model as an SIDM candidate.
Instead of parametrizing the model in terms of those listed in Eq.~\eqref{eq:parameters}, we opt for the quantities $R$, $\alpha$, and $m_H$; furthermore, since Eq.~\eqref{eq:splitting} relates $\Ehf/E_0$ to $\alpha$ and $R$, we may use $\Ehf/E_0$ in place of $R$ as a free parameter.
We consider two specific cases for the hyperfine splitting, $\Ehf = 10^{-4} E_0$ and $\Ehf = 10^{-5} E_0$, and determine what regions of the remaining parameter space can address the small-scale structure puzzles.
These values are chosen such that the cooling from inelastic scatterings may lead to interesting observable effects in dwarf-scale to cluster-scale halos.
Larger values of $\Ehf\gtrsim 10^{-3}E_0$ require correspondingly larger relative particle velocities in order to upscatter to the hyperfine excited state; for $\Ehf=10^{-3}$ this means that cooling effects would only occur in halos with $\vrms > 260~\km/\s$.
If the hyperfine splitting $\Ehf$ is decreased below $10^{-6}E_0$, the energy losses from upscatterings become negligible and our collisional cross sections approach those obtained in the elastic approximation (see Appendix~\ref{sec:elastic}).
Splittings of $\Ehf\sim 10^{-4}$ are of particular interest to us as they lead to large swaths of $\mH$--$\alpha$ parameter space in which collisional heating effects can solve the aforementioned small-scale structure puzzles.

In the following analysis, we work under the simplifying assumptions that the dark matter halos are completely neutral and are not affected by excitations to the $n=2$ state.
For the reasons that we discuss later on, the effects of excitations to the $n=2$ (see Sec.~\ref{subsec:cooling}) state as well as collisional ionizations (see Sec.~\ref{subsec:ionization}) may become non-negligible at galaxy cluster scales for $\Ehf=10^{-5}E_0$, the smaller of the two hyperfine splittings considered here.
Without a more detailed treatment of the dark atomic physics, it is possible that for hyperfine splittings $\sim 10^{-5}E_0$ these effects may change the cluster-scale halo structure in such a way as to become inconsistent with current observations.
To be conservative, one should not interpret the results below for $\Ehf=10^{-5}E_0$ as predictions, but should instead use them as a comparison to the more straightforward $\Ehf=10^{-4}E_0$ case to see how the results are affected for different hyperfine splittings.

\subsection{Cosmological considerations}
\label{subsec:cosmology}

Our goal is to uncover the cosmological phenomenology of neutral atomic dark matter in which the dominant inelastic scattering mode is through dark hyperfine transitions.
We must first map out the region of parameter space that is allowed for this model.

Under what circumstances does dark matter today consist of dark hydrogen bound states, with no dark ions remaining?
In analogy with SM hydrogen, there is a period of dark recombination in the early universe, and for large regions of parameter space, the majority of dark ions do indeed form into a neutral bound atomic state.
We assume the Universe has no overall dark charge, so perfect recombination would result in no remaining dark ions.
Most of the recombination occurs in the range $0.007 \lesssim T_D/B_H \lesssim 0.01$, where $T_D$ is the temperature of the dark radiation bath~\cite{CyrRacine:2012fz}.
At the end of dark recombination, the residual ionization fraction $\chi_e$ is given by~\cite{CyrRacine:2012fz}
\begin{equation}
  \chi_e \sim 2\times10^{-16} \frac{\xi}{\alpha^6}
  \left(\frac{0.11}{\Omega_\textrm{DM} h^2}\right)
  \left(\frac{m_{\rm H}}{\GeV}\right)   \left(\frac{B_H}{\keV}\right) \ ,
  \label{eq:chi-e}
\end{equation}
where $\xi\equiv (T_{D,\textrm{L}}/T_{V,\textrm{L}})$ is the ratio of the dark radiation temperature to the visible-sector CMB temperature in the present-day late universe.
The number density of dark matter particles changes with $m_H$, where larger masses correspond to lower number densities and hence a later recombination redshift and a larger ionization fraction.
A higher binding energy, $E_0/2$, results in a larger ionization fraction because recombination is less efficient.
However, the dependence on $\alpha$ is much stronger: if $\alpha$ is too small, the interaction between ions is simply not strong enough for them to attract one another and form bound states.

Constraints in the $\mH$-$\alpha$ plane are plotted in Fig.~\ref{fig:m_alpha_constraints} below for different values of $\chi_e$.
Postrecombination, we will require $\chi_e \lesssim 0.01$.
With this simplification, we assume all the dark matter is in the form of dark hydrogen and do not consider processes such as dark ion-ion or ion-hydrogen scattering.
Furthermore, we avoid dark particles with long-range forces, which affect structure formation~\cite{Kaplan:2009de,Kaplan:2011yj}.
Note that dark hydrogen typically recombines to the $n=2$ state and not the $n=1$ state analyzed in Sec.~\ref{sec:model}.
However, the lifetime of the $n=2$ state is very short ($\ll 1$ year~\cite{CyrRacine:2012fz}), so we expect all dark hydrogen to have settled into its $n=1$ ground state by the time of structure formation.

A complication for our desired dark hydrogen model is that dark hydrogen can potentially form molecules and affect halo cooling.
The formation of dark molecular hydrogen H$_2$ may occur through (1) neutral-neutral dark atom processes $\mathrm{H+H}\rightarrow\mathrm{H_2}+\gamma$ or (2) processes requiring a dark electron or proton catalyst ($\mathrm{e^- + H}\rightarrow\mathrm{H^- +\gamma},~ \mathrm{H^- +H}\rightarrow \mathrm{H_2+e^-}$), ($\mathrm{H^+ +H}\rightarrow\mathrm{H_2^+ + \gamma, H_2^+ + H}\rightarrow\mathrm{H_2 +H^+}$), where $\gamma$ is the dark photon.
The first type of process is very suppressed due to the fact that it must occur through a quadrupole transition~\cite{Hirasawa:1969db}.
The second type of process requires a free ionized population and thus may occur before or during recombination; however, at these times there are enough dark Lyman-Werner photons to photodissociate the dark H$_2$.
Hence, although a very small amount of dark molecular formation is possible, we do not consider it further.
However, in the visible sector, we know that even small traces of SM molecular hydrogen dramatically affect the cooling of gas in the first dark matter minihalos: for high-redshift ($z=23$) minihalos of masses $5\times10^5$--$10^6\msolar$, a SM molecular hydrogen fraction of $\lesssim10^{-3}$ can cool the innermost regions and precipitate gravitational collapse~\cite{Greif:2008qqa}.
Analogously, even a small amount of dark H$_2$ present may allow for much more efficient cooling in halos from excitations of the rotational and vibrational modes.
A comprehensive treatment of dark H$_2$ formation would be necessary to investigate this effect.

Finally, we consider the abundance and temperature of the dark sector, as both can dramatically affect the expansion history of the Universe and clustering of dark matter.
We assume there is a dark matter--antimatter asymmetry~\cite{Kaplan:2011yj}, and the full annihilation of dark antiparticles in the early universe yields the correct relic abundance of dark particles.
Since we assume the dark sector is overall charge neutral (i.e., an equal number of dark protons and electrons) by the time of recombination, the abundance is controlled by the heavier dark protons.
Thus, $\alpha$ is bounded from below by requiring that the dark antiprotons annihilate efficiently.
For $p + \bar{p} \rightarrow 2\gamma $ annihilation in the nonrelativistic regime, the velocity-averaged annihilation cross section is given by
\begin{equation}
  \langle \sigma v \rangle = \frac{\pi \alpha^2}{ m_p^2}
  = 3.66\times10^{-25} \cm^3/\s \left(\frac{\alpha}{0.01} \right)^2
  \left( \frac{m_p}{100~\GeV} \right)^{-2} \ .
\end{equation}
For Dirac dark matter, the thermal relic annihilation cross section is $\langle \sigma v \rangle \approx 4.4\times10^{-26} \cm^3/\s$~\cite{Steigman:2012nb}. Thus, for efficient annihilation of antiprotons, we have the following lower bound on $\alpha$:
\begin{equation}
 \alpha > 0.0035 \left(\frac{m_p}{100~\GeV} \right) \ .
\label{eqn:pp_constraint}
\end{equation}
We also need to include the annihilation to the hidden electrons and positrons, which would weaken the lower limit.
However, this is not required because these limits are much less constraining than our lower limits on $\alpha$ derived by imposing $\chi_e < 0.01$ on the late-time ionization fraction.
We will also find that in most of the parameter space where the SIDM phenomenology could be relevant for the small-scale puzzles, the $\alpha$ value will be larger than the lower limit in Eq.~\eqref{eqn:pp_constraint}.
Given these constraints, the dark matter abundance must be set by a dark matter--antimatter asymmetry.

Dark photons contribute to the effective number of relativistic degrees of freedom $N_\textrm{eff}$, measured at the time of last scattering and during big bang nucleosynthesis (BBN).
Since the dark sector is secluded from the visible sector, the temperatures in each are naturally allowed to differ.
If $\xi\lesssim 0.65$ (at $\sim1\sigma$), then we avoid BBN bounds on $N_\textrm{eff}$ for the range of $m_H, \alpha$, and $\Ehf$ considered here~\cite{CyrRacine:2012fz}.
One may attempt to motivate a natural value for $\xi$ by allowing the visible and dark sectors to interact, for instance, via a kinetic mixing term $\frac{1}{2}\epsilon_{k}F'_{\mu\nu}F^{\mu\nu}$, which would give the SM electrons a charge of $\epsilon_k e$ under the hidden U(1).
Then, the two sectors could come into thermal equilibrium through the process $e_\textrm{SM} +\gamma_{SM}\longleftrightarrow e_\textrm{SM}+\gamma$ at some temperature $T$ if the condition $T^2/M_\mathrm{Pl} = \alpha_\textrm{SM} \alpha \epsilon_k^2 T$ is met in the thermal bath~\cite{Hooper:2012cw}.
However, direct detection constraints from the LUX experiment place strong constraints on the mixing parameter $\epsilon_k \lesssim 2\times10^{-10}$ for the preferred regions of parameter space determined below~\cite{delNobile:2015uua}.
For the range of $\epsilon_k$ small enough to satisfy direct detection constraints, the condition for achieving thermal equilibrium is not reached prior to the freeze-out of SM $e^+ e^-$ annihilation.
The resulting low SM electron density causes the equilibration process between the dark and visible sectors to be inefficient, so the sectors do not achieve thermal equilibrium; thus, we do not have a well-motivated value to assume for the ratio $\xi$ of their present-day temperatures.
If initially set by inflationary reheating, the visible and dark sectors could have different temperatures depending on the inflaton couplings to the respective sectors.
We use the value $\xi=0.6$ in the following work. By using a value close to the upper limit on $\xi$, the contours we show later on in Sec.~\ref{subsec:minimum_halo_masses} may be interpreted as approximate upper limits on the minimum halo mass.

In summary, our requirement that dark matter consists exclusively of dark hydrogen bound states means that we only consider $\chi_e < 0.01$.
In order not to exceed the tight BBN constraints on the light degrees of freedom in the early universe, we require that the temperature of the dark sector be no greater than approximately 0.65 times the temperature of the visible sector.
Both of these constraints are easily met with a secluded dark U(1) sector.

\subsection{Cross sections, lifetimes, and structure formation}
\label{subsec:viscosity}

One of the difficulties in determining the effects of SIDM models on structure formation is that it is unclear how the microphysical scattering can be represented by macroscopic simulation particles in $N$-body experiments or in more general macroscopic descriptions of halos.
In this section, we advocate the use of the viscosity cross section of $n=1$ ground-state dark hydrogen atoms to model the microphysics of atomic dark matter in the evolution of dark matter halos.
We first show that nearly all dark atoms should be in their hyperfine ground state, and then motivate our choice of the viscosity cross section to model the scattering-induced energy flow in a dark matter halo.

Dark hydrogen may be in either its hyperfine ground or excited state.
For the hyperfine splittings we consider here, the time scale in Eq.~\eqref{eq:decay-width} for decays from the hyperfine excited state to the ground state is always much less than the time scale for excitation via upscattering.
Specifically, for parameter space of interest in Sec.~\ref{subsec:elastic_constraints}, the excited state lifetime is always $\Gamma^{-1}\ll 1~\Gyr$; a particle that upscatters into the excited state almost always emits a dark photon and returns to the ground state before scattering with another particle.
Hence, we focus on cross sections in which \textit{both particles are initially in the ground state}.

A dark matter halo can be altered if interactions occur that transfer momentum: elastic scatterings between atoms allow for heat to flow into the cold halo interior and increase the velocity dispersion of the inner halo relative to the CDM case.
The momentum-transfer cross section $\sigma_{\textrm{tot},T}$ (introduced in Sec.~\ref{subsec:scattering}) is commonly used in the literature to describe astrophysical constraints on the SIDM cross section, since it suppresses the far-forward scattering case of $\theta = 0$, which is equivalent to no interaction occurring.
However, $\sigma_{\textrm{tot},T}$ preferentially weights backward scattering, which also does not change the velocity distribution away from the CDM case, despite the fact that a large amount of momentum is transferred.
An alternative is instead to use the viscosity cross section $\sigma_{\textrm{tot},V}$ to favor scattering in the transverse direction.
While this choice may not be the fully correct quantity to use for comparison with SIDM constraints from simulations, we argue that it better captures the relevant SIDM physics at lowest order compared to the momentum-transfer cross section.
Additionally, the viscosity cross section is more physically well motivated to use with identical particles, for which forward and backward scattering are identical, and we agree with Ref.~\cite{Cline:2013pca} that the use of $\sigma_{\textrm{tot},T} = \sigma_\textrm{tot}$ for identical particles~\cite{Goldberg:1986nk,Krstic:1999lr,Jamieson:2000yg,CyrRacine:2012fz} is unwarranted.
Although the overall structure of the halo should be insensitive to the quantum mechanical nature of individual dark matter particles, it is reassuring that the viscosity cross section provides a consistent description for SIDM limits on both the macroscopic and the microscopic scales.
Thus, we consider $\sigma_{gg}$, $\sigma_{++}$, and their viscosity counterparts.
Recall from Sec.~\ref{subsec:scattering} that a single excitation from the scattering of two ground-state atoms does not occur because of selection rules (for $aa \to ab, ad$) or because total angular momentum is not conserved at a fixed orbital angular momentum (for $aa \to ac$).
Instead of the total $\sigma_{\textrm{tot},V}$, it is
\begin{equation}
  \sigma_V \equiv \sigma_{gg,V} + \sigma_{++,V}
  \label{eq:sigV}
\end{equation}
that we use to compare with target SIDM cross sections in Sec.~\ref{subsec:elastic_constraints}.

The use of the viscosity cross section is complicated by introducing a hyperfine splitting, and $\sigma_{\textrm{tot},V}$ may not be an appropriate quantity to use with SIDM either.
Keep in mind that the constraints from SIDM simulations assume elastic scattering---kinetic energy is conserved.
There may be regions of parameter space where $\sigma_{++,V}$ is a substantial contribution to $\sigma_V$ at energies near the hyperfine threshold, resulting in a significant loss in kinetic energy.
In this case, applying SIDM constraints using $\sigma_V$ is not necessarily a valid comparison, and we discuss this issue further in Sec.~\ref{sec:inelastic}.

\begin{figure}[t]
  \centering
  \includegraphics[scale=0.5]{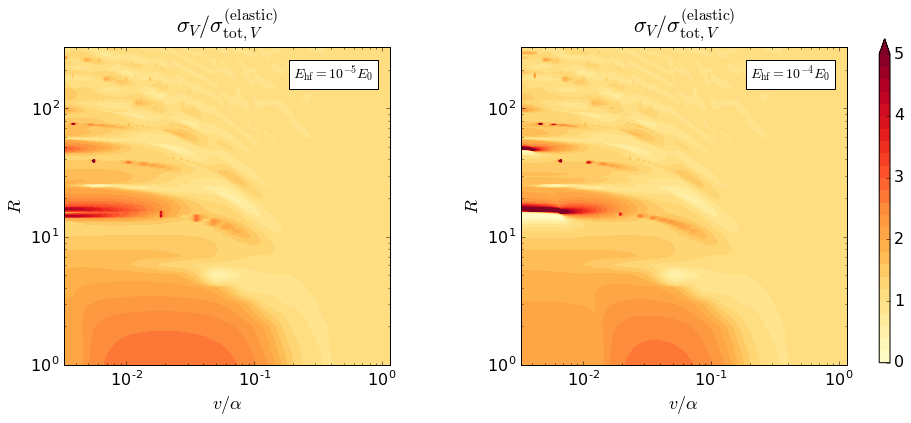}
  \caption{Parameter space scan comparing the viscosity cross section \eqref{eq:sigV}, which involves only hyperfine ground-state atoms, to the total viscosity cross section in the elastic approximation, used in Ref.~\cite{Cline:2013pca}.
    The left plot shows $\sigma_V$ with $\Ehf = 10^{-5}E_0$, while the right plot shows $\sigma_V$ with $\Ehf = 10^{-4}E_0$.
    The spin averaging factors for $\sigma_V$ and $\sigma_{\textrm{tot},V}^{(\textrm{elastic})}$ are different, so we do not necessarily expect $\sigma_V$ to be less than $\sigma_{\textrm{tot},V}^{(\textrm{elastic})}$.}
  \label{fig:sigVcompare}
\end{figure}

Figure \ref{fig:sigVcompare} shows the ratio of $\sigma_V$ (with $\Ehf = 10^{-4}E_0$ and $\Ehf = 10^{-5} E_0$) to $\sigma_{\textrm{tot},V}^{(\textrm{elastic})}$ (with $\Ehf = 0$) and demonstrates how the inclusion of the hyperfine interaction affects the cross section, as compared to Ref.~\cite{Cline:2013pca}.
Since we assume all dark atoms are in their hyperfine ground state, the spin-averaging factor for $\sigma_V$ is unity; meanwhile, $\sigma_{\textrm{tot},V}^{(\textrm{elastic})}$ contains more contributions from the other scattering channels, but it has a spin-averaging factor of $1/16$.
Thus, in our comparison of $\sigma_V$ and $\sigma_{\textrm{tot},V}^{(\textrm{elastic})}$, it is not generically true that $\sigma_V$ is the strictly smaller quantity.
At higher energies, the two cross sections are comparable, so using either results in very similar regions of acceptable SIDM parameter space.
At lower energies, there is a resonant effect from the scattering of low-velocity particles; these particles can exchange multiple dark photons and form quasibound states, resulting in an enhancement of the scattering cross section.
In this regime, $\sigma_V$ can be larger or smaller than the total elastic viscosity cross section by a factor of a few, and the resulting shapes of the acceptable SIDM regions will differ.
Although the SIDM constraints might look similar between our model and its counterpart in the elastic approximation, the crucial difference comes from the potentially significant energy losses that result from hyperfine transitions.
We discuss this issue further in Sec.~\ref{sec:inelastic}.

\subsection{SIDM halo profiles}
\label{subsec:elastic_constraints}

We now show how we use the viscosity cross section to identify interesting regions of dark atom parameter space.
Self-interacting dark matter models are motivated by their ability to produce cored density profiles in dark matter halos below $\sim 10^{11} \msolar$ (corresponding to $\vrms\lesssim 100~\km/\s$), thereby relieving tensions between the predictions for small-scale structure from collisionless $N$-body simulations and the inferred halo profiles of observed galaxies~\cite{Rocha:2012jg,Elbert:2014bma}.
Simulations of galaxies in this mass range find that hard-sphere scattering cross sections of $\sigma/m \sim 0.5$--$5~\cm^2/\g$ are capable of reproducing the cored density profiles of low surface-brightness galaxies (and perhaps dwarf spheroidals)~\cite{Rocha:2012jg,Elbert:2014bma}.
We thus require that the atomic dark matter models in our allowed region of parameter space result in cross sections $\sigVisc/\mH$ within this range for halos characterized by velocities of $\vrms=30$--$100~\km/\s$, which roughly corresponds to halo masses of $5\times10^9$--$10^{11}~\msolar$.

One may also calculate a target range for the velocity-dependent viscosity cross section at higher velocities of $\vrms\sim 1000~\km/\s$ using the core sizes of cluster-mass halos.
The scattering cross section required to produce a core of radius $r_\mathrm{c}$ may be approximated by assuming that the size of the cluster core is equal to the maximum radius at which the average dark matter particle will scatter at least once during the lifetime of the halo,
\begin{equation}
\label{eqn:scattering_time}
 \textrm{t}_\textrm{scatter}(r_\mathrm{c}) =
 \mathrm{t}_\mathrm{age} =
 \left( \sqrt{\frac{16}{3 \pi}} \frac{\sigVisc(v)}{\mH} \rho(r_\mathrm{c})~\vrms(r_\mathrm{c}) \right)^{-1} \ ,
\end{equation}
where $\rho(r)$ is the dark matter density at radius $r$.
The 1D velocity dispersion in a halo $\vrms$ and the average relative collisional velocity $v$ between particles are related by $v\approx\sqrt{2} \vrms$.
Reference \cite{Kaplinghat:2015aga} sets the cluster age $\mathrm{t}_\mathrm{age}=5$ Gyr and uses the halo profiles reported for the set of relaxed galaxy clusters in Ref.~\cite{Newman:2012nw} to derive the cluster-scale cross sections; they find that the observed core sizes may be reproduced if the cross section is $\sim0.1~\cm^2/\g$ at cluster velocities $\vrms\sim 1000~\km/\s$.
We use the inferred cluster-scale cross sections, velocities, and uncertainties from Ref.~\cite{Kaplinghat:2015aga} and require that our atomic dark matter models must have a viscosity cross section within this range at velocities $\vrms\sim 1000~\km/\s$.

While target cross sections for velocity-dependent SIDM may be obtained using the core sizes in dark matter halos across a wide range of characteristic velocities, upper limits on the scattering cross section at high velocities may be derived from observations of cluster-scale systems.
Constraints on SIDM cross sections at larger scales may be derived from measurements of merging galaxy clusters~\cite{Dawson:2011kf, Harvey:2015hha}, displacements of galaxies from cluster centers long after merging~\cite{Kim:2016ujt}, and ellipticities of dark matter halos~\cite{Buote:2002wd, Richard:2009yd}.
A constraint on $\sigVisc/\mH\lesssim1~\cm^2/\g$ at the cluster scale may be derived from the observed ellipticities of cluster halo profiles inferred through gravitational lensing~\cite{Richard:2009yd, Peter:2012jh}: if the SIDM cross section is too high, then repeated scatterings of particles in cluster halos will transform the halo shape from a triaxial ellipsoid into a sphere.
We use this value as an upper limit on $\sigVisc/\mH$ at cluster scales in the analysis below.

Another constraint on self-interaction cross sections may be derived from the lack of observed deceleration of the dark matter components due to a drag force in systems of merging clusters, although for a completely different type of cross section than can be produced in atomic dark matter models.
Reference \cite{Harvey:2015hha} uses observations of multiple merging cluster systems to place an upper limit of $\sigma/m<0.47~\cm^2/\g$ at relative velocities $v\sim 900~\km/\s$.
We note that this limit is not a true constraint on our parameter space, as it assumes an SIDM model where anisotropic and frequent scattering events with low momentum exchange give rise to a drag force with a $v^2$ dependence within merging clusters.

At the high end of the galactic mass scale, past works~\cite{CyrRacine:2012fz, Feng:2009hw} have also used the halo ellipticity of the elliptical galaxy NGC~720 as inferred through x ray observations~\cite{Buote:2002wd} to set an upper bound on the scattering cross section.
However, we do not use NGC~720 to set an upper limit at its velocity scale $\vrms\sim 250~\km/\s$ for the following reasons~\cite{Peter:2012jh}: (1) particles in such halos may on average undergo multiple scatterings over the halo lifetime while still preserving ellipticities, and (2) NGC~720 is the only object at this mass scale which has so far been measured to have an elliptical halo, and the scatter in simulated halo shapes may still allow for this halo to be accommodated as an outlier even if $\sigVisc/\mH \gtrsim 0.1~\cm^2/\g$.

\begin{figure}[t]
  \centering
  \includegraphics[width=3truein]{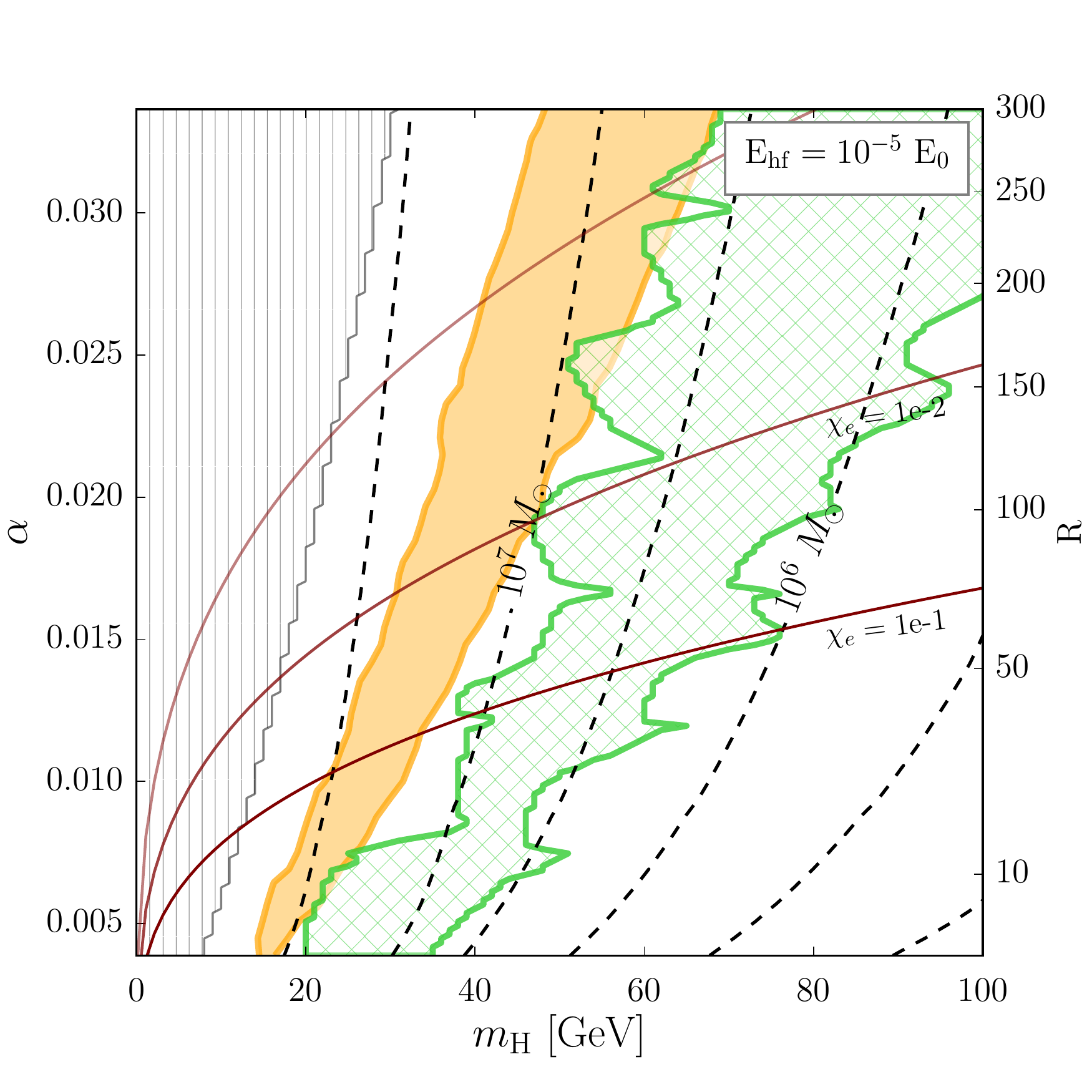} \hspace{1cm}
  \includegraphics[width=3truein]{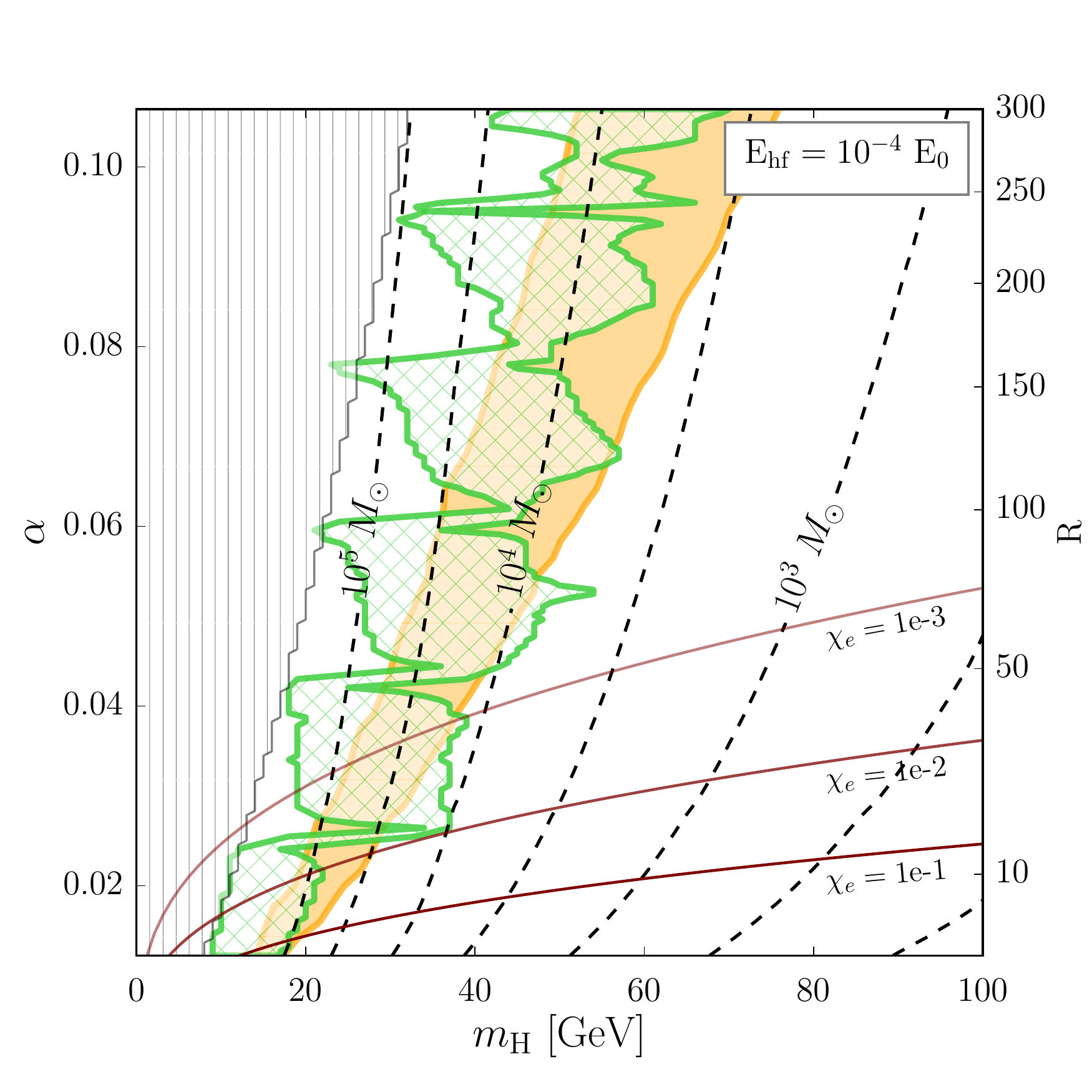}
  \caption{Viscosity scattering cross sections are calculated at the velocities of interest at each point in the $\mH$--$\alpha$ plane and then used to determine which areas either satisfy target cross sections or are in tension with observations.
    The hyperfine splitting is fixed to $\Ehf=10^{-5}~E_0$ in the left panel and $\Ehf=10^{-4}~E_0$ in the right panel.
    The vertical hatched grey area is disfavored by measurements of cluster halo ellipticities and corresponds to the region where $\sigVisc/\mH> 1~\cm^2/\g $ for velocities $\vrms=1000~\km/\s$.
    We show contours of the ionization fraction $\chi_e=10^{-1}$, $10^{-2}$, and $10^{-3}$; we consider $\chi_e \lesssim 10^{-2}$ sufficiently low to ignore dark ions, which excludes a large portion of the displayed parameter space for $\Ehf=10^{-5}~E_0$.
    Points within the cross-hatched green region satisfy $0.5~\cm^2/\g < \sigVisc/\mH < 5~\cm^2/\g $ for velocities $\vrms=30$--$100~\km/\s$, which approximates the condition for cores to form in lower-mass halos.
    Points within the solid orange region provide the best-fit viscosity cross sections for cores to form in relaxed cluster halos.
    The dashed lines show contours of constant minimum halo mass for values of $\Ehf, \mH,$ and $\alpha$ in our model, assuming $\xi = 0.6$.
    Lower, allowed values of $\xi$ lead to smaller minimum halo masses.}
  \label{fig:m_alpha_constraints}
\end{figure}

In Fig.~\ref{fig:m_alpha_constraints} we show the regions of $\mH$--$\alpha$ parameter space which satisfy the target cross section ranges for dwarf and low surface brightness galaxies, as well as galaxy clusters.
The cross-hatched green areas denote the region where $0.5~\cm^2/\g < \sigVisc/\mH < 5~\cm^2/\g $ in halos with $\vrms=30$--$100~\km/\s$, while the solid orange shaded areas denote the 68\% confidence limit for the best-fit region where the viscosity cross sections can reproduce cluster core sizes.
The resonant structure of the scattering cross section is evident in the green curves, since the associated velocities are quite low; the orange curves, on the other hand, do not probe the resonant regime.
The vertical hatched grey region is excluded by the cluster halo shape constraint of $\sigVisc/\mH< 1~\cm^2/\g $ in halos with $\vrms>1000~\km/\s$.
After imposing the constraint on $\sigVisc/\mH$ from cluster halo shapes, the regions of allowed parameter space in Fig.~\ref{fig:m_alpha_constraints} extend to lower masses ($\lesssim 100$ GeV) than those given in Refs.~\cite{CyrRacine:2012fz} and~\cite{Cline:2013pca}.
The overlap between the cross-hatched green region and the solid orange region encloses the values of $\mH$ and $\alpha$ which give the desired viscosity cross sections at both of the velocity scales of interest.
For $\Ehf = 10^{-5}~E_0$ there is very little overlap region, and much of the parameter space shown is excluded by our requirement that the Universe is neutral: $\chi_e \lesssim 0.01$.
For both cases, the calculation of the preferred SIDM regions in green and orange may not be reliable for $\chi_e > 0.01$.

\subsection{Minimum halo masses}
\label{subsec:minimum_halo_masses}

Prior to kinetic decoupling, the coupling between the dark radiation and the dark matter affects the growth of density perturbations and suppresses the small-scale matter power spectrum relative to predictions from CDM.
The dark radiation-matter coupling leads to dark sector analogs of the phenomena of diffusion (or Silk) damping and baryon acoustic oscillations.
The growth of density perturbations below the damping scale $r_\textrm{D}$ is damped, leading to a cutoff in the matter power spectrum at small scales.
For a detailed review and explanation of these effects, see Refs.~\cite{Bertschinger:2006nq,Feng:2009mn, CyrRacine:2012fz, CyrRacine:2013fsa,CyrRacine:2015ihg}.

One may associate a minimum dark halo mass $M_\mathrm{min}$ with the smallest perturbation mode below which the growth of structure is suppressed, given by~\cite{Bertschinger:2006nq}
\begin{equation}
 M_\mathrm{min} \simeq
 0.1 \left(\frac{T_\textrm{dec}}{\MeV}\right)^{-3}\msolar \ .
 \label{eqn:Mmin}
\end{equation}

The minimum halo mass is set by the decoupling temperature $T_\mathrm{dec}$, which in turn is set by the physics of the atomic dark matter.
Unlike the equivalent scenario involving SM hydrogen, the dark atoms do not necessarily become transparent to dark photons soon after recombination---for high enough values of $\alpha$, the contribution to the opacity from Rayleigh scattering between photons and neutral atoms may keep the dark plasma opaque even if it is not ionized~\cite{CyrRacine:2012fz}.
There are thus two expressions for $T_\mathrm{dec}$, depending on whether Compton or Rayleigh scattering provides the dominant contribution to the dark matter opacity prior to decoupling.
If Rayleigh scattering dominates, we use the following equation for $T_\mathrm{dec}$~\cite{CyrRacine:2012fz}:
\begin{equation}
 T_\mathrm{dec}^\mathrm{Rayleigh} \simeq 7\times10^{-4}  B_H \left[ \frac{1}{\alpha^6 \xi^3} \bigg(\frac{B_\mathrm{H}}{\keV} \bigg) \bigg(\frac{\mH}{\GeV} \bigg)  \right]^\frac{1}{5}~~,
 \label{eqn:rayleigh_tdec}
\end{equation}
where $\xi$ is the ratio of the dark radiation temperature to the visible sector CMB temperature in the present-day late universe.
As mentioned in Sec.~\ref{subsec:cosmology}, we take this ratio to be $\xi=0.6$ so that the minimum halo masses calculated are approximately upper limits.

If Compton scattering is the dominant source of opacity, we follow the method of Ref.~\cite{CyrRacine:2013fsa} to solve for the scale factor at decoupling $a_\mathrm{dec}$ and then convert this to the temperature $T_\mathrm{dec}$.
We approximate that the dark electrons and photons decouple when the expansion rate begins to exceed the Thomson scattering rate, i.e. when $H\simeq n_H \chi_e \sigma_\mathrm{Thomson}$.
This leads to the following equation for the scale factor at decoupling $a_\mathrm{dec}$:
\begin{equation}
 a_\mathrm{dec}^3 + \frac{\Omega_\mathrm{R}}{\Omega_\mathrm{m}} a_\mathrm{dec}^2 =
 \frac{1}{\Omega_\mathrm{m} h^2} \left[ \epsilon_D \alpha \xi
   \bigg( \frac{B_\mathrm{H}}{\mathrm{eV}} \bigg)^{-1}
   \bigg(\frac{\mH}{\mathrm{GeV}} \bigg)^{-\frac{1}{6}} \right]^2 \ .
\end{equation}
The constant $\epsilon_D$ is obtained by fitting to the numerically calculated ionization fraction and thermal evolution of the dark sector and is  approximately $\epsilon_D\sim8\times10^{-3}$ for the range of $\alpha$ considered here~\cite{CyrRacine:2013fsa}.
We find that the Rayleigh scattering case dominates for the parameter space considered here, and thus the minimum halo masses shown in Fig.~\ref{fig:m_alpha_constraints} are calculated using the decoupling temperatures given by Eq.~\eqref{eqn:rayleigh_tdec}.

We calculate the minimum halo masses in our region of $\mH$-$\alpha$ parameter space and show these as dashed contours in Fig.~\ref{fig:m_alpha_constraints}. In general, the minimum halo masses in the allowed region of parameter space lie below the current observational limits.
The highest minimum halo mass that is not ruled out by the constraints from cluster shapes is $M_\mathrm{min}\sim 10^{7.5}\msolar$ for hyperfine splittings of $\Ehf=10^{-5} E_0$, or $M_\mathrm{min}\sim 10^5~\msolar$ for $\Ehf=10^{-4} E_0$.
If the parameter space is constrained by demanding that velocity-dependent elastic cross sections produce cores at both high and low halo masses, then this would lead to a prediction of a minimum halo mass of around $M_\mathrm{min}\sim 10^{7}~\msolar$ for fixed $\Ehf=10^{-5}$, or $M_\mathrm{min}\sim 10^{3.5-5}~\msolar$ for fixed $\Ehf=10^{-4}$.
Halos of these sizes may be observed in next-generation substructure lensing or galactic tidal stream surveys~\cite{Hezaveh:2014aoa,Bovy:2016irg}.

The temperature ratio $\xi$ may take on different values in various inflationary reheating scenarios (as long as the BBN constraint of $\xi\lesssim 0.65$ is still satisfied).
From Eqs.~\eqref{eqn:Mmin} and \eqref{eqn:rayleigh_tdec}, $M_\mathrm{min}\varpropto \xi^{9/5}$ if $T_\mathrm{dec}$ is set by the dominance of Rayleigh scatterings: lower values of $\xi$ result in lower minimum halo masses.
For a sufficiently large difference in potential values of $\xi$ (e.g. 0.1 versus the upper limit of 0.65), the minimum halo mass can vary by an order of magnitude.
If one were able to observe the matter power spectrum cutoff associated with $M_\textrm{min}$ in an atomic dark matter scenario, this measurement could be translated into a lower bound on the dark to visible sector temperature ratio in various regions of the $(\Ehf,\mH,\alpha)$ parameter space.

In summary, we find regions of dark atom parameter space that satisfy observed constraints on the elastic SIDM cross section on scales from dwarfs to clusters.
Because dark acoustic oscillations lead to a dark-sector temperature-dependent truncation of the halo mass function on potentially observable scales, it may be possible to fully constrain the atomic dark matter model with halo core sizes and the halo power spectrum.

\section{Consequences of Inelastic Scattering}
\label{sec:inelastic}

The constraints on atomic dark matter so far come from assuming elastic scattering.
One of the most interesting aspects of dark atoms is that they have excited states, which admit inelastic processes.
In this section, we consider the magnitude of inelastic hyperfine scattering in our model.
The net effect of two particles upscattering and then decaying back to the ground state is an overall loss in kinetic energy; this provides a mechanism for cooling in dark matter halos which may potentially counterbalance or dominate over the heating mechanism provided by elastic scatterings.
Depending on the values of $\Ehf$, $\mH$, and $\alpha$, either one or both of these effects may have a large influence on the evolution of a halo. In the following discussion, we investigate whether inelastic cooling effects may significantly impact the halo structure in any regions of the parameter space considered in this work.

\subsection{Comparison of viscosity and upscattering cross sections}
\label{subsec:cross_sections}

\begin{figure}[t]
\centering
 \includegraphics[width=2.8truein]{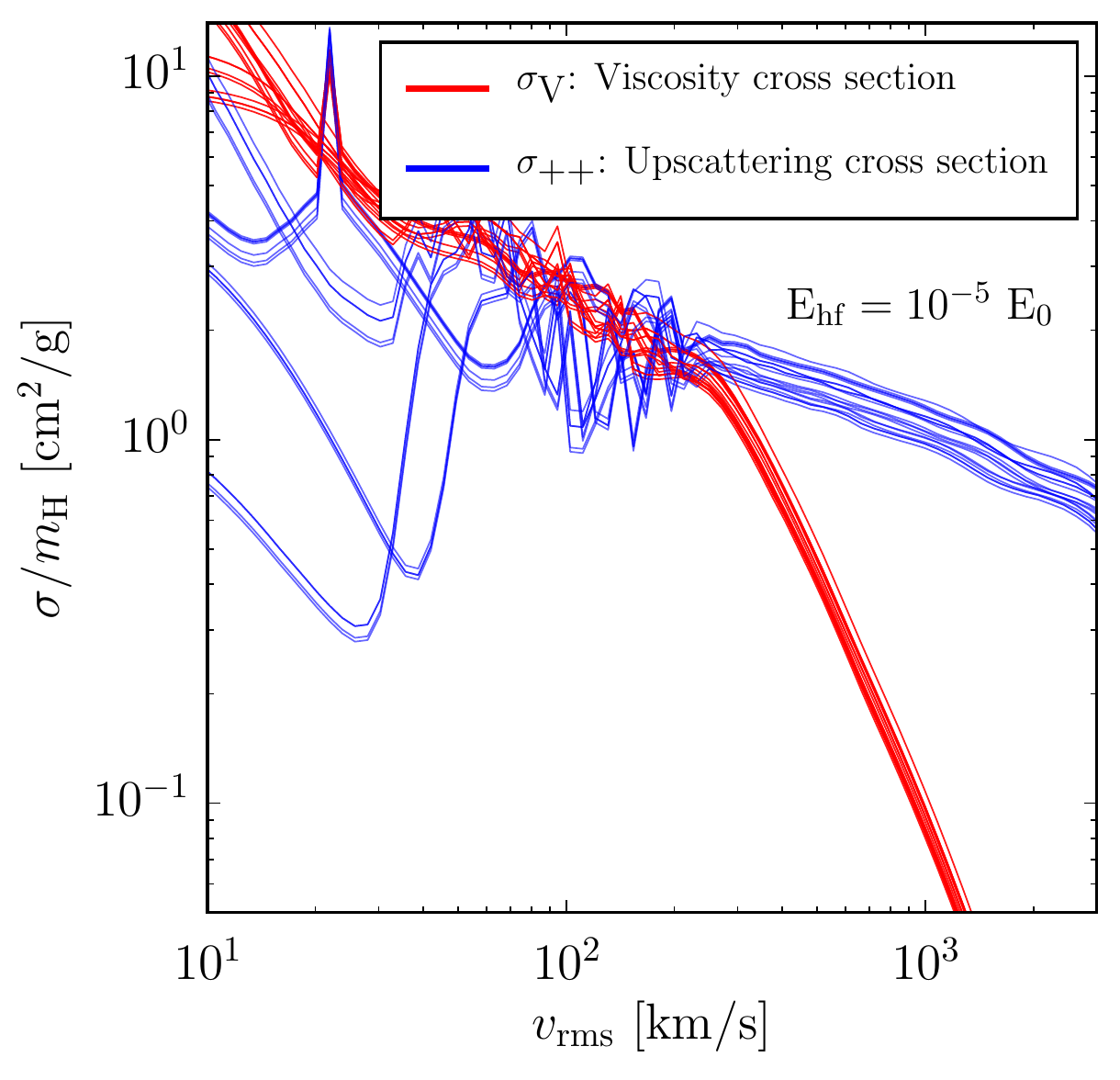}\hspace{0.5in}
  \includegraphics[width=2.8truein]{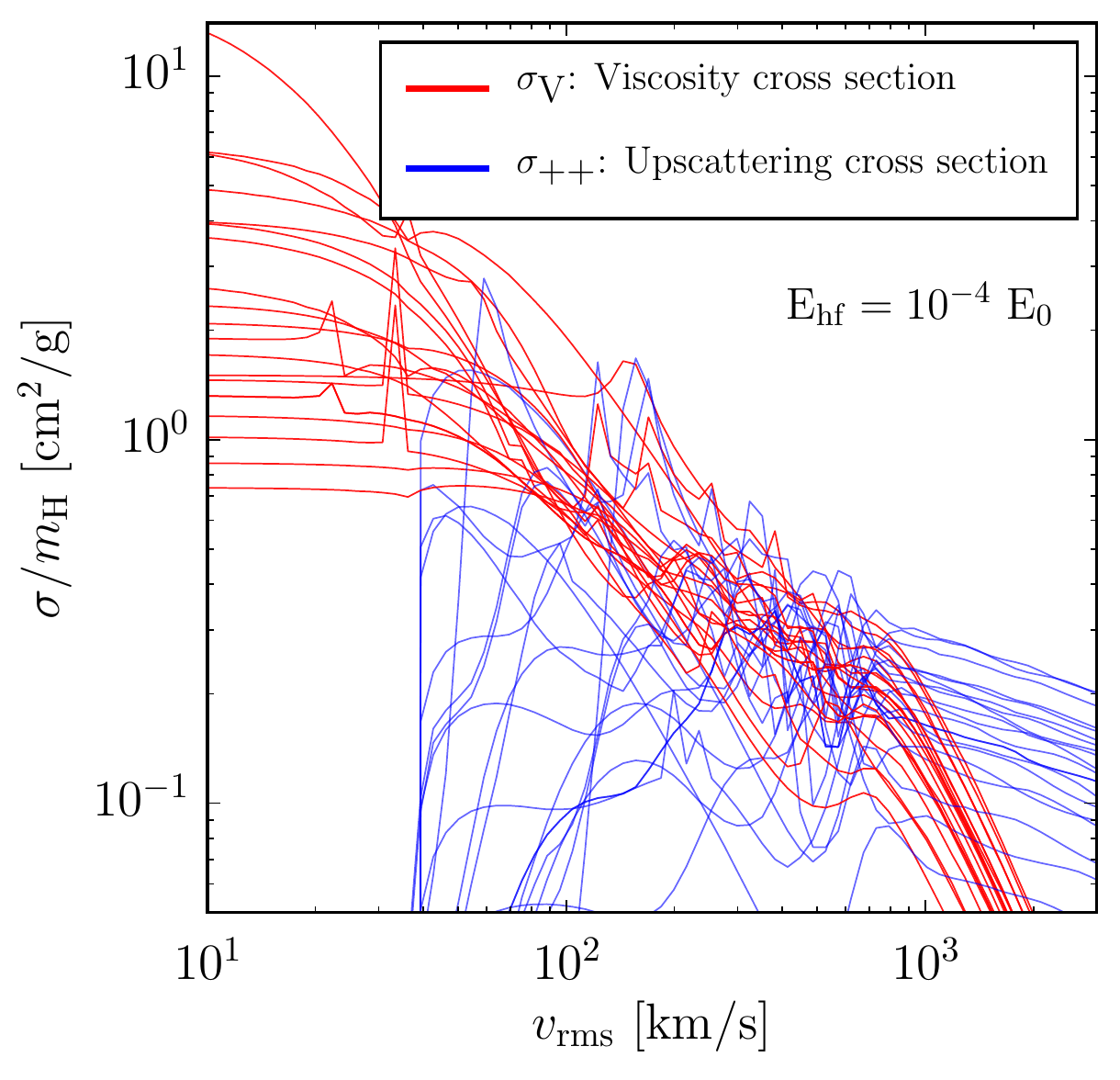}
  \caption{Viscosity (red curves) versus upscattering (blue curves) cross sections per unit mass as a function of halo velocity.
    Because of the short decay time scale for the hyperfine excited state, we assume that both initial particles are in the ground state.
    The value of the hyperfine splitting is fixed to $\Ehf=10^{-5}~E_0$ and $\Ehf=10^{-4}~E_0$ in the left and right panels, respectively.
    Each line is drawn randomly from values of $\mH$ and $\alpha$ lying within the overlapping regions of Fig.~\ref{fig:m_alpha_constraints} corresponding to parameters which lead to cores consistent with observations in both cluster-scale halos as well as dwarf- to LSB-scale halos.}
  \label{fig:sig_m_multiple}
\end{figure}

In Fig.~\ref{fig:sig_m_multiple} we plot examples of upscattering and viscosity cross sections per unit mass for values of $\mH$ and $\alpha$ lying within the allowed regions of parameter space in Fig.~\ref{fig:m_alpha_constraints} which may produce cores in cluster, LSB, and dwarf halos consistent with observations.
We plot these cross sections for multiple values of $\Ehf$, $\mH$, and $\alpha$ in order to show the range of velocity-dependent behavior allowed in our model which may resolve small-scale structure issues.
In the following discussion, we explore the phenomenology that may arise in different regions of parameter space due to the different velocity-dependent behaviors of the viscosity and upscattering cross sections.

We now compare the velocity dependence of the viscosity and upscattering cross sections at cluster-scale velocities $\vrms\sim 1000~\km/\s$.
At velocities of order $\mathcal{O}(100)~\km/\s$ and lower, the viscosity cross section is generally comparable to or larger than the upscattering cross section; at higher velocity scales above $1000~\km/\s$, the viscosity cross section decreases at a steeper rate such that it falls well below the upscattering cross section.
We choose to plot cross sections and velocities in units of astrophysical observables in Fig.~\ref{fig:sig_m_multiple}, though the same quantities can be plotted in terms of the geometric cross section $\pi a_0^2$ and atomic energy $E_0$.
The shape of the viscosity cross section changes noticeably at two velocities or energies; these changes are more easily seen for the case where $\Ehf = 10^{-4}E_0$.
At very low energies, the viscosity cross section is $s$ wave and constant.
There is a break in the viscosity cross section near $E = E_0 R^{-3/2}$, where the de Broglie wavelength of the atoms can probe the structure of the lowest-order Van der Waals interaction potential ($\sim 1/x^6$), and higher partial waves begin to contribute.
For $R\sim 100$, this break occurs near $v \sim 100~\km/\s$.
In atomic units, $\sigVisc/\pi a_0^2$ scales roughly as $(E/E_0)^{-0.4}$.
There is another power-law break in the viscosity cross section near $E \sim 0.1\, E_0$ or $v \sim 1000~\km/\s$ for $R\sim 100$.
At these higher energies, higher order multipoles of the Van der Waals potential are able to be probed, and thus higher partial waves contribute to the viscosity cross section.
The increased importance of higher partial waves cause the viscosity cross section to fall quickly as $\sigVisc/\pi a_0^2 \sim (E/E_0)^{-1.3}$.
This behavior allows for the model to produce high viscosity cross sections at dwarf scales that decrease quickly enough with energy to become consistent with cluster observations at higher velocities.

Depending on the values of $\Ehf$ and $\alpha$, the inelastic upscattering cross section surpasses the viscosity cross section in halos with characteristic velocities as low as $\vrms\sim 300~\km/\s$, and can be over an order of magnitude larger at cluster scales.
The ratio $\siginel/\sigVisc$ increases with decreasing $\Ehf$; for $\Ehf=10^{-5}$ the upscattering cross section is $\mathcal{O}(10)$ times higher than the viscosity cross section at cluster scales.
However, this does not necessarily mean that cooling is more efficient in atomic dark matter models with lower hyperfine splittings---although upscatterings may occur more frequently in these models, the smaller values of $\Ehf$ mean that less energy is lost when the particles upscatter into the excited state and decay.
We quantify the relative effectiveness of the heat transfer and energy loss mechanisms in the following Sec.~\ref{subsec:cooling}.

The right panel of Fig.~\ref{fig:sig_m_multiple} demonstrates an interesting feature of the atomic dark matter model: if the average kinetic energy of two incoming particles in a halo is lower than the hyperfine splitting $\Ehf$, then the dark atoms cannot upscatter to the hyperfine excited state and the inelastic cross section drops precipitously.
The value of $\Ehf$ thus sets a halo scale below which our mechanism for cooling is ``turned off.''
Dark atoms in halos with velocities below this scale may still be upscattered if they lie in the high-velocity tail of the velocity distribution, but the overall upscattering rate will be severely lowered by this effect.
For hyperfine splittings of $\Ehf=10^{-4}~E_0$, upscatterings are suppressed in halos with $\vrms\lesssim 40~\km/\s$.
For hyperfine splittings of $\Ehf=10^{-5}~E_0$, upscatterings are only suppressed for $\vrms\sim 1$--$2~\km/\s$, which corresponds to halos of mass $M_\textrm{halo}\sim 10^6~\msolar$.
If the cooling effects of collisional upscattering lead to observable effects in the structural evolution of the dark halo, then measurements of halo profiles below and above this turn-off velocity may allow us to infer a hyperfine splitting value in an atomic dark matter scenario.

\begin{figure}[t]
\centering
  \includegraphics[width=3.58truein]{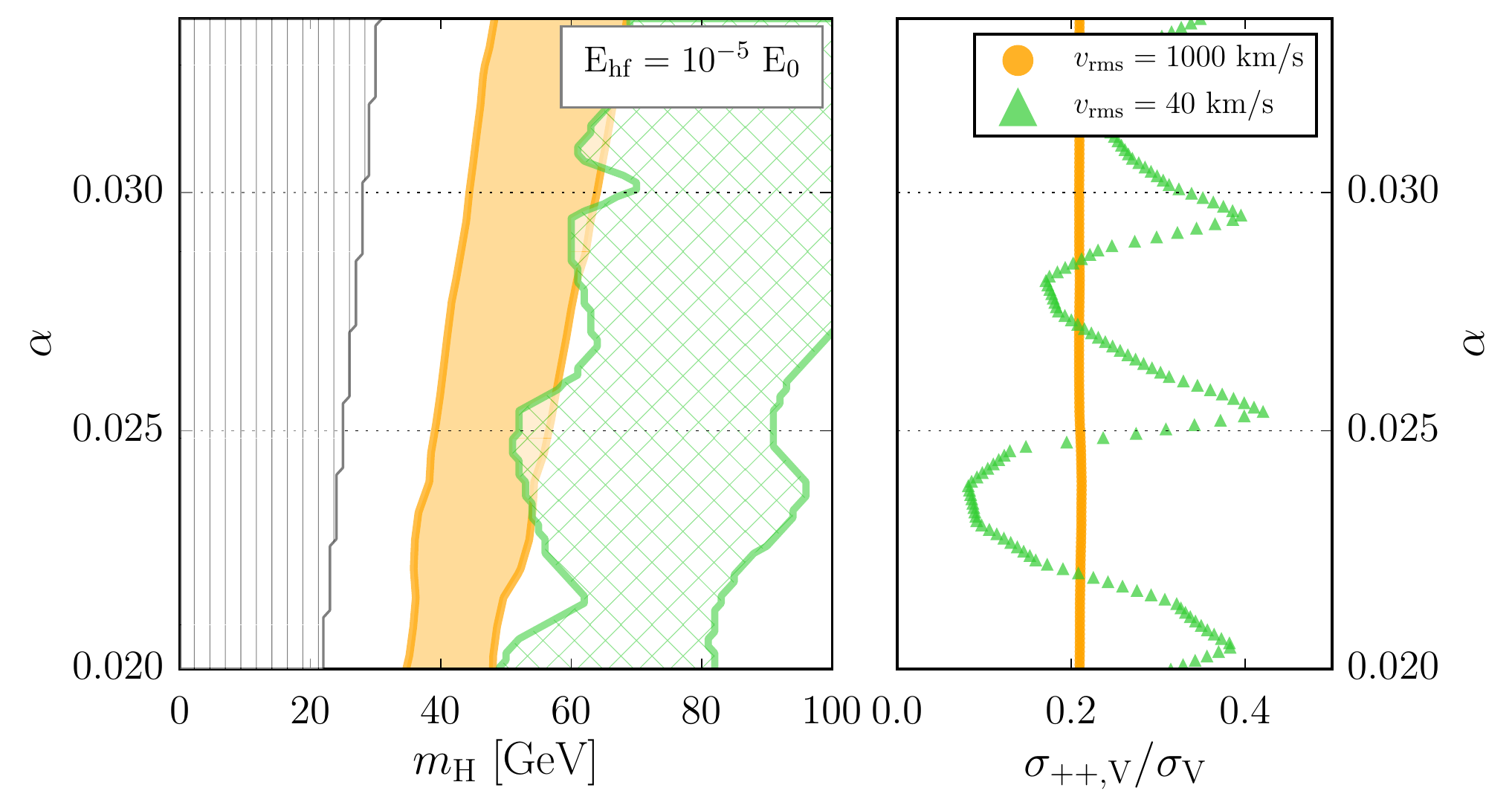}
  \includegraphics[width=3.5truein]{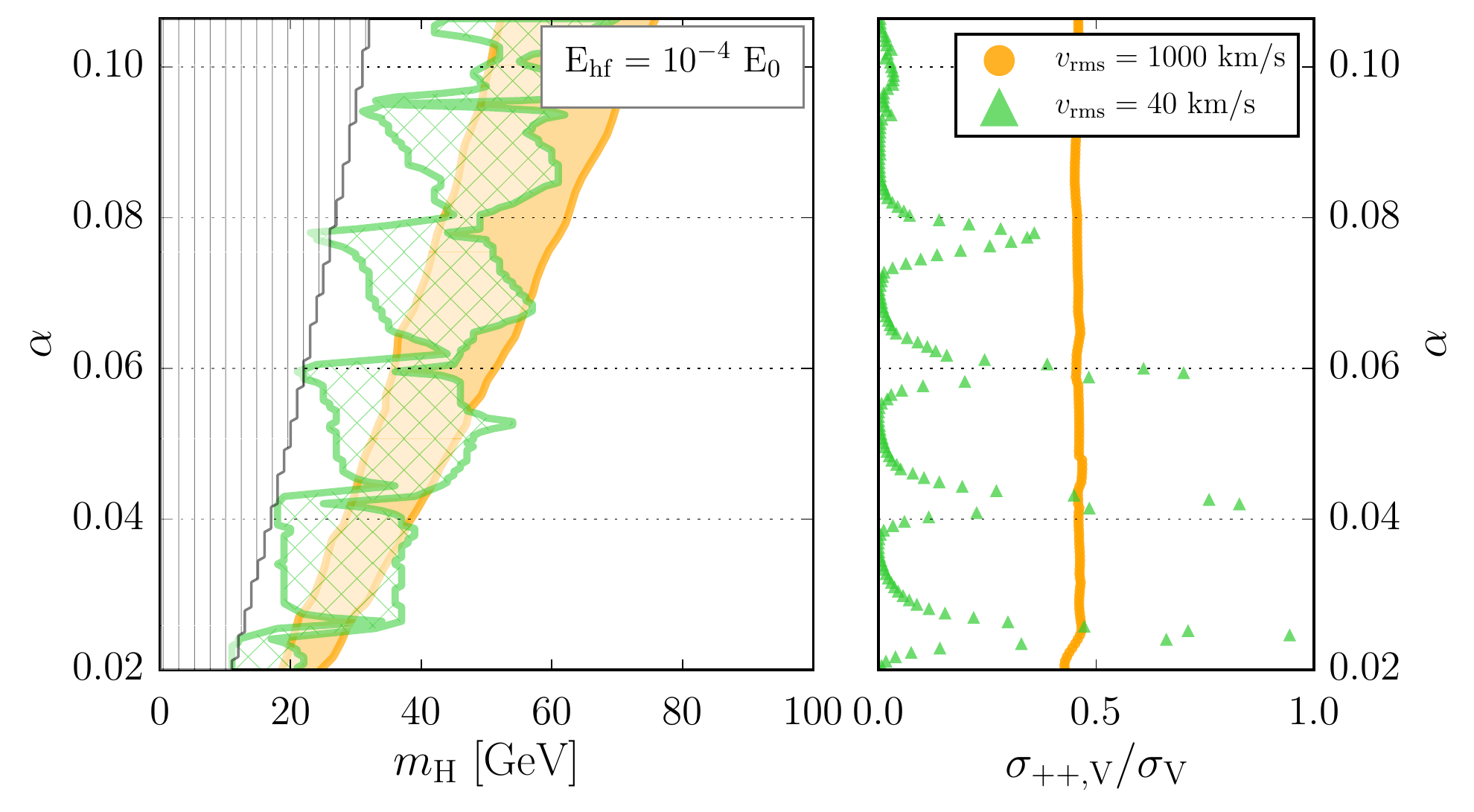}
  \caption{The left subpanels in each figure show the constrained (vertical grey hatched) and target (solid orange and cross-hatched green) areas of parameter space.
    The right subpanels show the ratio of the inelastic viscosity cross section to the viscosity cross section as a function of $\alpha$.
    The left and right figures are shown for fixed hyperfine splittings of $\Ehf=10^{-5}$ and $10^{-4}$, respectively.}
  \label{fig:sigratios_visc}
\end{figure}

In the right-hand subpanels of Fig.~\ref{fig:sigratios_visc}, we plot the fraction of the viscosity cross section $\sigVisc$ that arises from the inelastic viscosity cross section $\sigma_{++,V}$ at velocities of $\vrms=40~\km/\s$ and $\vrms=1000~\km/\s$.
We are particularly interested in the fraction $\sigma_{++,V}/\sigVisc$ for the larger hyperfine splitting of $\Ehf=10^{-4} E_0$ at low velocities $\vrms=40~\km/\s$: if this fraction is large, inelastic collisions are no longer approximately equivalent to elastic collisions in terms of momentum transfer, and comparison to existing SIDM constraints becomes difficult.
From the right panel of Fig.~\ref{fig:sigratios_visc}, we see that for the majority of the target parameter space in this case (green triangles), the inelastic viscosity cross section contributes only a small ($\lesssim 0.2$) fraction of the viscosity cross section at low velocities.
We therefore assume that our comparison to existing SIDM cross section constraints are valid for these values of $\alpha$ at which the viscosity cross section is close to the elastic viscosity cross section.
We do note that there are narrow ranges of $\alpha$ where the inelastic contribution to the viscosity cross section is significant ($\gtrsim 0.8$).
Nonetheless, we retain this definition for $\sigma_V$ for two reasons: we want to include $\sigma_{++,V}$ far from the hyperfine threshold where all scattering processes are approximately elastic, and it is not clear to what extent inelastic processes near threshold affect the SIDM constraints.

\subsection{Halo cooling}
\label{subsec:cooling}

We now quantify the energy loss rate due to hyperfine upscatterings.
In what follows, we assume that halos are optically thin to the dark photons emitted following decays from the excited state.
In the limit $E_\gamma \ll E_\mathrm{Ly\alpha}$, which is valid for $E_\gamma=\Ehf$, the Rayleigh scattering cross section for dark photons of energy $E_\gamma$ is approximately given by~\cite{CyrRacine:2012fz}
\begin{equation}
\sigma_\mathrm{Rayleigh} \approx \frac{81\pi}{24}\left(\frac{\alpha}{m_e}\right)^2 \left(\frac{E_\gamma}{E_\mathrm{Ly\alpha}}\right)^4 \ .
\end{equation}
Since this quantity is negligible for dark photons with energies equal to the hyperfine splittings considered here, we assume that the emitted photons free stream out of the halo after emission.

Although the upscattering cross section can be large relative to the total viscosity cross section, the more important quantity to compare is the energy flow from each process.
We now calculate the energy losses expected from inelastic upscatterings and compare this energy loss rate to the rate of inward heat flow due to dark atom-atom collisions.
We outline below the net kinetic energy loss per particle expected per hyperfine upscattering and decay.
We define the parameter
\begin{equation}
 \varepsilon\equiv\frac{m_{H^\ast}^2 - m_H^2}{4 m_H^2} = \frac{ \Ehf } {2m_H} \ ,
\end{equation}
where $m_{H^\ast} \equiv m_H + \Ehf$ and $m_H$ refer to the excited and ground state masses, respectively.
For the halos and parameter space studied here, $v_0\ll 1$ and $\varepsilon\ll 1$, where $v_0$ is the incoming relative velocity of the colliding particles.
In this limit, the net change in kinetic energy per particle per upscattering in the center-of-momentum frame is simply $\Delta \mathit{KE}_\textrm{upscatter}\approx-2m_H\varepsilon = -\Ehf$.

After the upscattered particle decays, its velocity in the \textit{lab} frame is given by
\begin{equation}
 v_f^2 = 1 - \frac{1-v_0^2}{1+4\varepsilon} \left( 1- \frac{2\varepsilon}{1+4\varepsilon} \frac{1-v_0^2}{(1-v_0 \cos{\theta})}  \right)^{-2}
\end{equation}
where $\theta$ is the angle between $\boldsymbol{v_0}$ and the outgoing dark photon.
The change in kinetic energy after undergoing this decay is
\begin{equation}
 \Delta\mathit{KE}_\textrm{decay} = 2 \varepsilon \mH
 \left(8\varepsilon +2v_0^2 - v_0 \cos{\theta}\right) \ .
\end{equation}
After averaging over possible angles $\theta$, this is a net \textit{increase}---emitting a dark photon imparts a net positive kick velocity to the final ground state atom.
However, the increase in kinetic energy from decay processes is $\mathcal{O}(\varepsilon^2)$ or $\mathcal{O}(\varepsilon v_0^2)$ (the dominant term depends on the value of $\Ehf$ and the halo in question), while the decrease from upscattering processes is $\mathcal{O}(\varepsilon)$.
[The net $\Delta\mathit{KE}_\textrm{upscatter}$ of both particles is still $\mathcal{O}(\varepsilon)$ after shifting back to the lab frame.]
Henceforth, we will approximate the change in kinetic energy per particle per upscattering as $\Delta\textit{KE}\approx \Ehf$ when investigating the regimes in which cooling effects become important.

The rate of energy loss in a thin shell of width $dr$ at a radius $r$ is given by
\begin{equation}
 4\pi r^2 dr \Gamma_\mathrm{upscatter} n_H \Delta E \approx
 4\pi r^2 dr \sqrt{\frac{16}{3\pi}}
 \frac{\siginel(v) \vrms(r) \rho^2(r) \Ehf }{\mH^2} \ ,
 \label{eqn:coolingrate}
\end{equation}
where we have used the above reasoning to assume that the average energy lost per particle upscattering is approximately equal to $\Ehf$.
The total amount of energy lost due to atomic upscatterings in a halo over its lifetime may be estimated by integrating Eq.~\eqref{eqn:coolingrate} over the radius $r$ and multiplying by the lifetime.
We verify that the energy lost due to collisional cooling is never more than 0.001 times the total initial kinetic energy of the halo---for the range of parameters studied here, hyperfine upscatterings cannot disrupt the entire halo.
However, as we demonstrate below, the energy losses from inelastic upscatterings can be up to 0.1--0.5 times the rate of inward heat flow from scatterings within the inner halo for particular ranges in $\alpha$.
Hyperfine upscatterings may therefore play an important role in the structural evolution of the inner halo if the atomic SIDM model parameters lead to significant cooling rates.

To calculate the rate of heat flow resulting from particle collisions, we treat the dark halo as a fluid with the luminosity $L$ at radius $r$ given by ~\cite{Balberg:2002ue, Pollack:2014rja}
\begin{equation}
 \frac{L}{4 \pi r^2} = -\kappa \frac{\partial T}{\partial r} =
 -\frac{3}{2} a b v \frac{\sigVisc}{\mH}
 \left[ a \left( \frac{\sigVisc}{\mH}\right)^2
   + \frac{b}{C} \frac{4\pi G}{\rho v^2}  \right]^{-1}
 \frac{\partial v^2}{\partial r} \ .
 \label{eqn:heattransfer}
\end{equation}
The dimensionless coefficients $a$ (which describes hard sphere scattering)%
\footnote{This value given for the coefficient $a$ in Eq.~\eqref{eqn:heattransfer} assumes elastic scatterings.
  As noted previously in Sec.~\ref{subsec:cross_sections}, $\sigVisc$ has contributions from both elastic and inelastic scattering cross sections.
  However, $\sigVisc$ can be considered as an approximately elastic cross section if 1) $\Ehf\ll \mH v^2$, i.e. the hyperfine splitting is small compared to the initial energies of the interacting particles, or 2) the viscosity upscattering cross section $\sigma_{++,V}$ does not contribute significantly to $\sigVisc$.
  Either one or both of these conditions are met for a large majority of our favored regions in parameter space (see Fig.~\ref{fig:sigratios_visc}).
  We therefore consider the use of this value for $a$ to be reasonable.}, $b$ (which describes the short mean-free-path regime), and $C$ (which describes the scale at which the transition between long- and short-mean-free path regimes occurs) are taken to be $a=\sqrt{16/\pi}, b=25\sqrt{\pi}/32$, and $C\approx0.75$ as in Ref.~\cite{Pollack:2014rja}.
In Sec.~\ref{subsec:viscosity}, we explain why the viscosity cross section $\sigVisc$---as opposed to the transfer cross section $\sigma_T$---is the quantity that best describes the rate of events which result in a net transfer of energy.
In line with this reasoning, we use our calculated values for $\sigVisc$ in Eq.~\eqref{eqn:heattransfer} when calculating the rate of heat flow.

\begin{figure}[t]
  \centering
  \includegraphics[width=3.3truein]{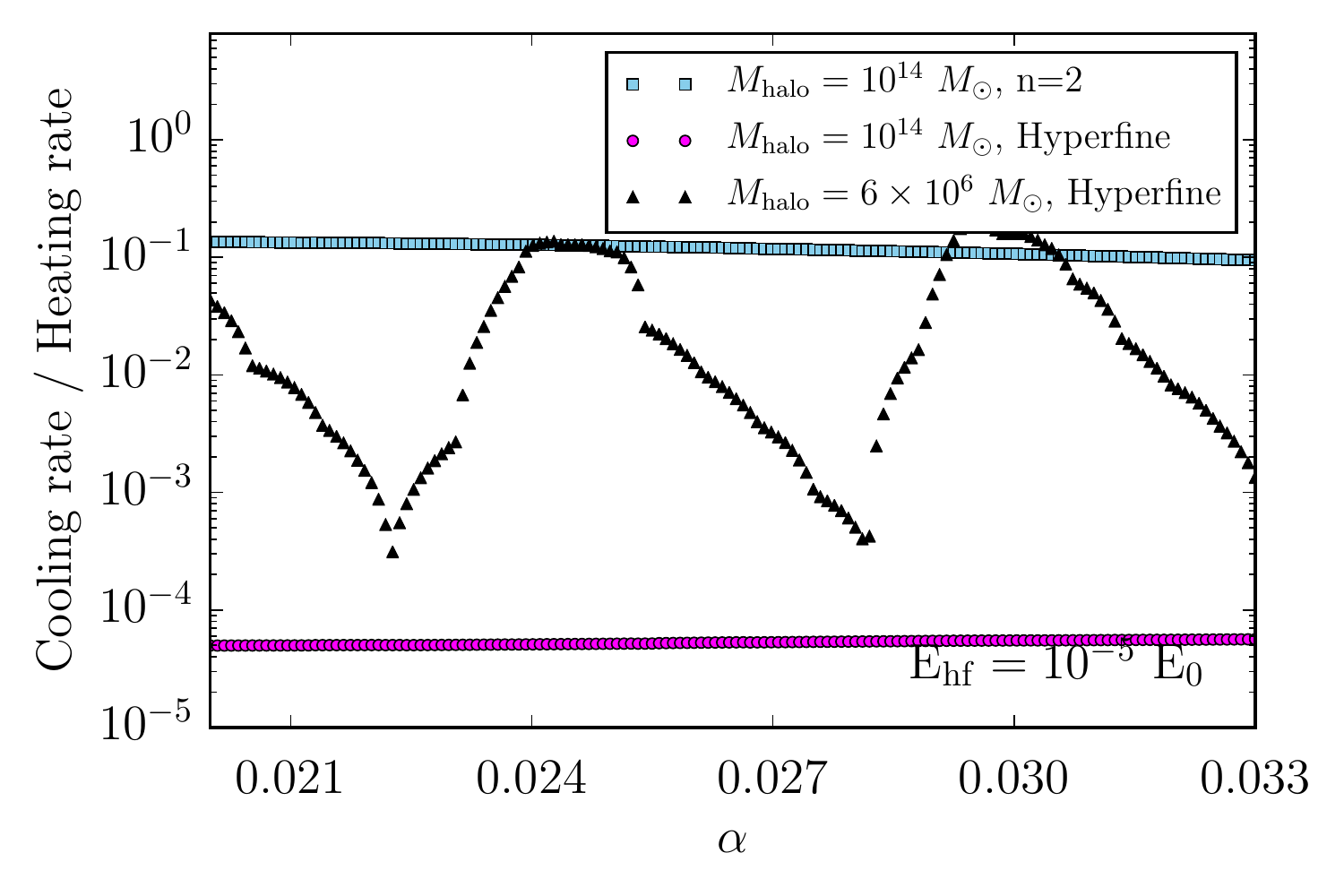} \hspace{0.2in}
  \includegraphics[width=3.3truein]{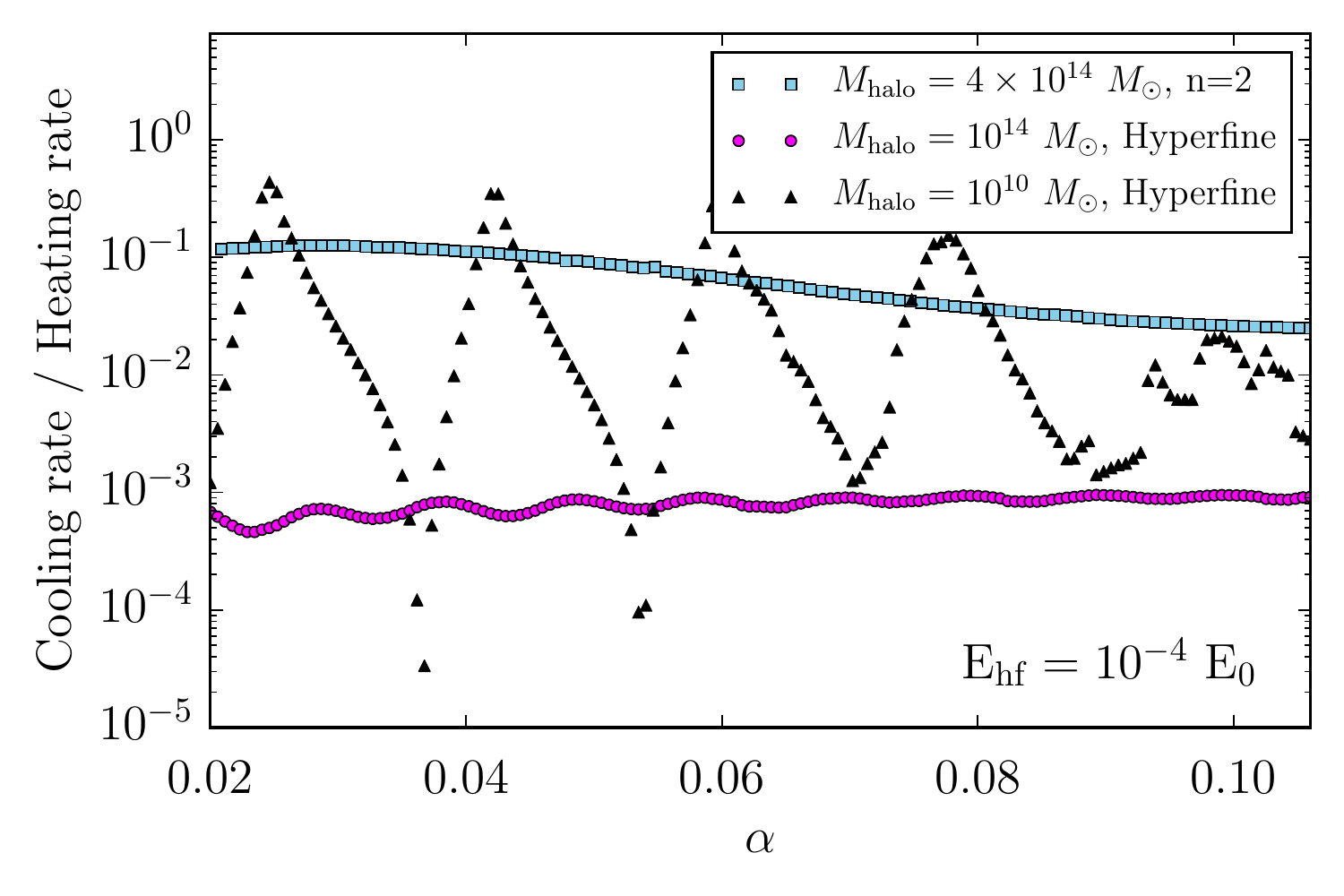} \\
  \caption{We compare the effects of SIDM heating and cooling mechanisms in atomic dark matter halos by plotting the ratios of outward energy flow lost through cooling over the inward heat flow from scatterings at a radius of $r=0.5\; r_s$ in the halo, which is approximately the radius at which the inward heat flow due to scatterings and the outward energy loss due to upscatterings are greatest, as well as $r=r_s$.
    The left and right figures in both rows are shown for fixed hyperfine splittings of $\Ehf=10^{-5}$ and $10^{-4}$, respectively.
    The lower mass halo plotted in each panel (black triangles) corresponds to the smallest halos in which upscatterings to the hyperfine excited state are not suppressed by low particle velocities.
    We also show the cooling to heating ratios at cluster scales for cooling through hyperfine excitations (magenta circles) and $n=2$ excitations (blue squares). See Sec.~\ref{subsec:n2} for details and discussion regarding our estimation of the $n=2$ cooling rate.}
  \label{fig:cooling_heating_rates}
\end{figure}

In Fig.~\ref{fig:cooling_heating_rates} we show the ratio of heat lost through upscattering and decays to heat inflowing through collisional processes in a thin shell at radius $r=0.5\; r_s$ in low-mass and high-mass halos.
The low-mass halo is chosen to be approximately the lowest-mass halo in which the inelastic upscattering rate is not suppressed by the average particle velocity being lower than the hyperfine splitting.
For a hyperfine splitting of $10^{-4}E_0$ ($10^{-5}E_0$), this corresponds to a halo mass of $10^{10}~\msolar$ ($6\times10^{6}~\msolar$).
The high-mass halo corresponds to a cluster-scale halo with $M_\textrm{halo}=10^{14}~\msolar$. We choose to plot this ratio at the radius $r=0.5\; r_s$ as this is roughly where the cooling and heating rates are both maximized in the halo.

We find that cooling is preferentially important for small halos relative to big halos.
This is because the energy loss $\Delta \mathit{KE} \sim E_{\text{hf}}$ is fixed, while the typical kinetic energy per particle increases with increasing halo mass.
Furthermore, the cooling and heating processes have different overall effects on the halo: heating the inner part of the halo is caused by a transferral of energy within the halo, whereas the energy emitted as dark photons is presumably not reabsorbed and is instead permanently lost from the halo.
In Fig.~\ref{fig:cooling_heating_rates}, the cases with high cooling rates have high heating rates as well, so the moderate cooling-to-heating ratios could be underestimating the overall importance of cooling.

The structural evolution of the halo in instances of non-negligible cooling effects is nontrivial and may be modeled using numerical integration methods.
Evolving an atomic dark matter halo over cosmic time with the inclusion of dark cooling as well as the baryonic potential in the innermost region $r\lesssim 0.1\; r_s$ is beyond the scope of this paper, but will be addressed in future work.

\section{Additional Considerations at the Cluster Scale}
\label{sec:clusters}

In our above treatment of the interactions between neutral dark atoms, we do not consider the possibility of collisional ionizations or excitations to $n\geq2$ states.
This simplification is adequate for lower-mass halos in which dark matter particles have enough energy to excite the hyperfine state, but not enough energy to ionize or excite the $n=2$ state through collisions.
As halo masses increase and the typical particle velocities surpass the ionization and $n=2$ excitation thresholds, these processes may potentially affect the halo structure.
Below, we discuss the potential for these processes to affect cluster-scale halos.

\subsection{Upscatterings to the $n=2$ excited state}
\label{subsec:n2}

Above a particular halo mass scale, particle velocities may be high enough to collisionally excite atoms into the $n=2$ excited state, which would quickly decay back to the ground $n=1$ state.
This additional cooling mechanism may affect the halo structure if the relative particle velocities are above the threshold for upscattering one of the incident particles into the $n=2$ state,
\begin{equation}
  v^2 > \frac{9}{8 g_e g_p} \frac{\Ehf}{E_0} \ .
\end{equation}
For hyperfine splittings of $\Ehf=10^{-5}E_0$, relative particle velocities $v\gtrsim 500~\km/\s$ (corresponding roughly to a halo mass of $\sim10^{13}\msolar$) may result in upscattering to the $n=2$ state.
Relative velocities above $\sim 1600~\km/\s$ are needed for hyperfine splittings of $\Ehf=10^{-4}E_0$, which may be reached in massive clusters ($M_\mathrm{halo}\gtrsim4\times10^{14}\msolar$) or systems of merging clusters.

We use the following method to obtain approximate values for this cross section in order to estimate the potential cooling losses from $n=2$ upscatterings.
From the analytic derivation of cross sections for collisions between neutral ground-state SM hydrogen atoms presented in Ref.~\cite{Bates:1953lr}, we see that the $n=2$ upscattering cross sections $\sigma_{n=2}(\mathrm{1s+1s}\rightarrow\mathrm{1s+2s/2p})$ may be written using $v/\alpha$ as the independent variable, with $\sigma_{n=2}$ in units of the geometric cross section $\pi a_0^2$.
We then scale the experimental measurements of this cross section \cite{Hvelplund:1982fk, Barnett:1990qy} to estimate the collisional $n=2$ upscattering cross sections for the dark hydrogen analogs.

Using this scaling, the $n=2$ upscattering cross sections are typically much smaller than the hyperfine upscattering cross sections ($\sigma_\mathrm{n=2}\lesssim0.01~\sigma_{++}$).%
\footnote{Cross sections quoted in Secs.~\ref{subsec:n2} and \ref{subsec:ionization} make the simplifying assumption that all particle pairs have the same typical relative velocity for their position in the halo.
  If the cross section is appropriately averaged over the Maxwell-Boltzmann distribution and relative velocities below the minimum threshold for the interaction are excluded, results are consistent within $\sim20\%$.}
However, the energy lost per upscattering is much greater than [of order $(E_0/\Ehf)^{-1}$ times] the energy lost per hyperfine upscattering.
Since the $n=2$ cooling rate may thus be non-negligible, we estimate it using the expression on the left-hand side of Eq.~\eqref{eqn:coolingrate} with $\Gamma_\mathrm{upscatter}=n_H \sigma_\mathrm{n=2} v$ and $\Delta E= \Delta E_\mathrm{Ly\alpha} = 3/4\; B_H$.
We show this estimate of the $n=2$ cooling rate over the heating rate in Fig.~\ref{fig:cooling_heating_rates}.

For hyperfine splittings of $\Ehf=10^{-5}E_0$, we find that the $n=2$ cooling rate over the collisional heating rate can be up to $\sim0.1$ in cluster-scale halos of mass $10^{14}\msolar$.
Although the relative particle velocities in smaller halos are above the threshold for $n=2$ upscattering, the cross section $\sigma_{n=2}$ for these interactions decreases with velocity in this regime such that this ratio is an order of magnitude lower for a $10^{13}\msolar$ halo than for a $10^{14}\msolar$ halo.
For hyperfine splittings of $\Ehf=10^{-4}E_0$, halos must be at least $\sim4\times10^{14}\msolar$ in mass for enough particles to surpass the threshold velocity for $n=2$ upscattering. For a $4\times10^{14}\msolar$ halo, we find that the cooling-to-heating ratio from $n=2$ upscattering is $\simeq0.1$ for $\alpha\lesssim0.04$ and decreases to $\simeq0.02$ for $\alpha\simeq0.1$.
We therefore expect that while the cooling effects from $n=2$ upscattering processes may be large enough to affect halo structure, they do not affect the evolution and growth of lower mass halos and only become significant at the cluster scale.

\subsection{Ionization in the late universe}
\label{subsec:ionization}

Once halos form, dark hydrogen remains intact if there is insufficient energy in the system for ionization: $(v/\alpha)^2 < 1/f(R)$.
The particle velocities in a halo may be high enough to ionize the majority of dark atoms if the following condition is met:
\begin{equation}
  v^2 > \frac{3}{2 g_e g_p} \frac{\Ehf}{E_0} \ .
\end{equation}
For the hyperfine splitting $\Ehf=10^{-5}(10^{-4})E_0$, the above relation is satisfied for relative velocities of $v\gtrsim580 (1800)~\km/\s$: atoms have enough energy for ionization in isolated cluster halos if $\Ehf\sim10^{-5}$, or in merging clusters if $\Ehf\sim10^{-4}$. This raises the concern that dark matter halos above these velocities may contain a significant ionized component.
However, the above condition is necessary but not sufficient to ionize the majority of the dark atoms in a halo---the cross section for collisional ionization $\sigma_{i}$ must also be high enough to allow for particles to experience such an interaction over the cluster lifetime or merger time.

In a similar manner in which we use the analytic expression from Ref.~\cite{Bates:1953lr} to estimate the $n=2$ excitation cross sections, we scale the experimental measurements of the collisional ionization cross section~\cite{Hvelplund:1982fk, Barnett:1990qy} to estimate the collisional ionization cross sections for atomic dark hydrogen.
At its maximum, this cross section is approximately the geometric cross section: $\sigma_{i,\textrm{max}}\simeq \pi a_0^2$.
Thus, for hyperfine splittings of $\Ehf=10^{-4}$, the collisional ionization cross section is always $\sigma_i\lesssim 0.01~\cm^2/\g$, and we do not expect cluster halos to be significantly ionized.

However, the geometric cross sections in our preferred region of parameter space for splittings of $\Ehf=10^{-5}E_0$ are large enough ($\pi a_0^2\sim0.2$) such that the collisional ionization cross section may be as large as $\sigma_i \sim0.1~\cm^2/\g$ for relative velocities above $v\sim 2000~\km/\s$.
Massive clusters or systems of merging clusters above these velocities may become ionized if the hyperfine splitting is of order $\Ehf\sim10^{-5}E_0$.
Ionization may result in increased mass loss during mergers, cooling effects (due to recombination followed by emission of a photon), and a variety of possible scattering cross sections between ions, electrons, and atoms.
The complex effects of ionization on the structural evolution of a halo are not included in the comparison of our results with existing cluster-scale observations, and we caution that hyperfine splittings of $\Ehf=10^{-5}E_0$ in this model may alter the dark matter structure at high mass scales to be inconsistent with observations.
Again, for the aforementioned reason of low ionization cross section, this issue of late-time ionization does not significantly affect our results for splittings of $\Ehf=10^{-4}$.

\section{Conclusions}
\label{sec:conclusions}

In this paper, we have investigated a model of self-interacting dark matter that mimics the properties of atomic hydrogen.
Dark matter in the late universe takes the form of dark hydrogen, which is neutral under a new U(1) gauge force.
We do not assume a specific interaction between this new U(1) and the SM for the predictions in this paper.
The key features of our work are the inclusion of a hyperfine interaction, which induces an energy splitting in the ground state of dark hydrogen, and the calculation of the basic heat transport properties in halos, which allows us to identify the viable regions of parameter space where the small-scale puzzles can be solved.

Collisions of dark atoms in halos may induce hyperfine excitations, which then decay by emitting dark photons.
Halo cooling from this upscattering and subsequent energy loss works against halo heating that occurs from the scattering processes.
To study these effects on halo structure, we calculated the cross sections for dark hydrogen scattering over a wide range of parameter space, using techniques from standard hydrogen to aid in numerically solving the Schr\"{o}dinger equation.
The velocity dependence of the cross sections allows the heating and cooling mechanisms to operate differently on scales of dwarf spheroidal galaxies ($\vrms = 40~\km/\s$) compared to scales of galaxy clusters ($\vrms = 1000~\km/\s$).

We argue that the viscosity cross section where both the forward and backward scattering are suppressed is the better quantity, compared to the momentum-transfer cross section, to use when comparing to SIDM simulation results and observational constraints.
The velocity dependence of the viscosity cross section shows a sharp drop for kinetic energies larger than about $0.1\; E_0 \simeq 0.1\; \alpha^2 m_H/R$ as contributions from higher partial waves become important.
This allows the model to be consistent with cluster constraints. The typical cross section at $E=0.1\; E_0$ is roughly $10\; a_0^2$ and scales approximately as $E^{-1.3}$ above these energies.
For kinetic energies below $0.1\; E_0$, we see a steady increase in the viscosity cross section with decreasing relative velocity, which implies that the scattering processes are very important in small halos.
The viscosity cross section in this regime scales roughly as $E^{-0.4}$.

We have found regions of parameter space for the atomic dark matter model in which dark matter self-interactions can explain the measured core sizes in both dwarfs and clusters, while being consistent with all other observations including cluster halo shapes.
The solutions are not fine-tuned; for a hyperfine splitting that is about $10^{-4}E_0$, we find that much of the parameter space with $\chi_e < 0.01$ and dark hydrogen mass in the 10--100 GeV range is viable.
In this part of parameter space, the dark matter is in atomic form and we find that cooling mechanisms are generically important for the structure of low-mass halos (masses below $10^{10} M_\odot$) but not important enough to completely disrupt these halos.
An immediate consequence of this observation is that the collapse of small halos at early times will be affected by the cooling and, therefore, it is likely that the growth of the seeds of supermassive black holes will also be altered.
We leave this discussion for another paper.

The kinetic energy of dark matter particles in galaxy clusters is large enough to allow for additional atomic physics.
We find that collisional excitations to $n=2$ and ionizations could be significant processes in galaxy clusters for $\Ehf = 10^{-5}$.
For $\Ehf = 10^{-4}$, we show that the cooling rate due to these processes is subdominant to the heating rate and our predictions, which assume negligible scattering to $n=2$ and fully atomic dark hydrogen, are robust.
Thus, galaxy clusters are important astrophysical laboratories for testing atomic dark matter models.

The interactions between the dark matter and the light mediator in the early Universe modifies the kinetic decoupling of the dark matter.
The kinetic decoupling temperature may be used to estimate the minimum halo mass in the universe.
Assuming that the ratio of the hidden sector temperature to the visible photon temperature at late times is 0.6 (close to the maximum allowed by BBN constraints), we find that the range of halo minimum masses in the viable regions of parameter space are between $10^{3.5}$ and $10^7 M_\odot$.
These minimum masses are smaller than the host masses of the currently observed dwarf galaxies, but much larger than the minimum masses predicted for dark matter in weak-scale theories.
If the ratio of the temperatures is smaller (due to the fact that the two sectors were reheated to different temperatures and remained decoupled), then the minimum halo masses will be lower by a factor of $(\xi/0.6)^{9/5}$.

In summary, we have shown that an analog of hydrogen in the hidden sector is a viable self-interacting dark matter candidate that can alleviate the small-scale structure formation puzzles, and the dissipative nature of atomic dark matter provides a phenomenologically rich foundation to make observational predictions.

\section*{Acknowledgments}

We thank Chris Hirata, Kris Sigurdson, Xerxes Tata, and Jesse Thaler for useful discussions.
We used the LSODA software from LLNL~\cite{Hindmarsh:1983ode,Radhakrishnan:1993ode} to numerically solve the Schr\"{o}dinger equation.
All numerical calculations of cross sections were performed using the Ohio Supercomputer Center~\cite{OhioSupercomputerCenter1987}.
K.B. is funded in part by NSF CAREER Grant No. PHY-1250573.
A.K. is supported by NSF GRFP Grant No. DGE-1321846.

\appendix
\section{Numerical Work}
\label{sec:numerical}

\subsection{System of coupled Schr\"{o}dinger equations}
\label{sec:full}

To solve the system of Schr\"{o}dinger equations \eqref{eq:SE}, we work in the $\ket{F_A M_A F_B M_B}$ basis, and each partial wave solution occupies a particular point in the parameter space $(\mathcal{E},\Ehf/E_0,R)$.
Our numerical solver begins with a set of initial conditions at $x_i>0$ and finds the solution $\mathbf{F}_l(x)$ and its derivative at a sufficiently large value $x_f$.

Since the goal is to determine the scattering cross section from one channel in Table \ref{tab:channels} to another, we set the initial wave function and its derivative at $x_i$ to begin only in a particular channel.
There are 16 such choices, which correspond to 16 linearly independent solutions of Eq.~\eqref{eq:SE}.
We assign these solutions as the column vectors of a $16 \times 16$ solution matrix $\mathbb{F}_l$, and we order them such that $\mathbb{F}_l(x_i)$ and $\mathbb{F}'_l(x_i)$ are diagonal.
As $x\to 0$, the angular momentum term in the Schr\"{o}dinger equation dominates, and the analytic form of the wave function in this limit is known.
Thus, we set the initial condition of the $j$th component of the $j$th solution to be $[\mathbb{F}_l(x_i)]_{jj}=x_i$ and $[\mathbb{F}'_l(x_i)]_{jj}=(l+1)$ (with all other terms zero); the overall normalization is irrelevant.
As the numerical solver evolves to $x>x_i$, off-diagonal terms in $\mathbb{F}_l(x)$ appear, indicating that inelastic scattering into other channels has occurred.

At $x_f$ we match to the asymptotic solution
\begin{equation}
  \lim_{x\to\infty} \mathbb{F}_l =
  \mathbb{J}_l(kx) - \mathbb{N}_l(kx) \mathbb{K}_l \ ,
  \label{eq:asymptotic}
\end{equation}
where $\mathbb{K}_l$ is the reaction matrix.
$\mathbb{J}_l(kx)$ and $\mathbb{N}_l(kx)$ are diagonal matrices
\begin{align}
  [\mathbb{J}_l(kx)]_{ij} &=
  \begin{cases}
    \delta_{ij}\, k_i x\, j_l(k_i x) & \textrm{for } k_i^2 >0 \\
    \delta_{ij}\, k_i x\, \iota_l(k_i x) & \textrm{for } k_i^2 <0
  \end{cases}
  \\
  [\mathbb{N}_l(kx)]_{ij} &=
  \begin{cases}
    -\delta_{ij}\, k_i x\, n_l(k_i x) & \textrm{for } k_i^2 >0 \\
    -\delta_{ij}\, k_i x\, \kappa_l(k_i x) & \textrm{for } k_i^2 <0
  \end{cases}
\end{align}
where $j_l(kx)$ and $n_l(kx)$ are the spherical Bessel functions of the first and second kinds, and $\iota_l(kx)$ and $\kappa_l(kx)$ are the modified spherical Bessel functions of the first and second kinds.
If the wave number as defined in Eq.~\eqref{eq:wavenumber} is imaginary, then the channel is closed and is omitted from the $S$ matrix.
Note that the asymptotic matching would be different in the $\ket{SM_S IM_I}$ basis, because there are finite off-diagonal terms in the total potential at infinity.
By inverting Eq.~\eqref{eq:asymptotic}, we find
\begin{equation}
  \mathbb{K}_l =
  \left[\mathbb{Y}_l(x_f)\mathbb{N}_l(kx_f)-\mathbb{N}^\prime_l(kx_f)\right]^{-1}
  \left[\mathbb{Y}_l(x_f)\mathbb{J}_l(kx_f)-\mathbb{J}^\prime_l(kx_f)\right] \ ,
\end{equation}
where
\begin{equation}
  \mathbb{Y}_l(x) = \mathbb{F}^\prime_l(x) \left[\mathbb{F}_l(x)\right]^{-1} \ .
\end{equation}
The primes denote derivatives with respect to $x$, not $kx$.
The $j$th diagonal element of $\mathbb{K}_l(x)$ is proportional to $\tan \delta^{(j)}_l$, where $\delta^{(j)}_l$ is the partial wave phase shift associated with elastic scattering in the $j$th channel.

Since the range $[x_i,x_f]$ is finite, we must ensure that it yields a convergent expression for the phase shifts and is sufficiently independent of the choice for $x_i$ and $x_f$.
We make an initial guess by setting $x_i$ at the threshold where the angular momentum term begins to dominate over other terms and setting $x_f$ at the threshold where the angular momentum term and $k^2$ begin to dominate over the potential term $f(R)V(x)$.
With our beginning range $[x_i,x_f]$, we increase $x_f$ by 1\% until the phase shifts converge to 1\%.
We reset $x_f$ and repeat this process while decreasing $x_i$ by 10\% until the phase shifts converge to 1\%.
Once we have a reliable $\mathbb{K}_l$ matrix, we define the scattering $S$ matrix and amplitude as
\begin{align}
  \mathbb{S}_l &= (\mathds{1}+i\mathbb{K}_l)^{-1} (\mathds{1}-i\mathbb{K}_l) \\
  \mathbb{T}_l &= \mathds{1} - \mathbb{S}_l \ .
\end{align}
Finally, the cross section from the state $\ket{j}\equiv\ket{F_A M_A F_B M_B}$ to $\ket{i}\equiv\ket{F'_A M'_A F'_B M'_B}$ is found from summing over partial waves~\cite{Zygelman:2005lr} and is given by Eq.~\eqref{eq:cross-section}.
We truncate the sum over partial waves when they contribute less than 1\% to the cumulative cross section.

Although we have presented the procedure for solving \eqref{eq:SE} in terms of 16 coupled differential equations, the block diagonal form of $\mathbb{V}+\mathbb{W}$ allows us to break the problem into four sets of coupled equations, two of which are identical.
Even with this division, calculating the cross section over a large region of parameter space requires significant computational resources.
For energies much larger than threshold, we switch to an elastic approximation.

\subsection{Elastic approximation}
\label{sec:elastic}

In the limit of $\Ehf=0$, it is easiest to work in the basis $\ket{SM_S I M_I}$, where the potential $\mathbb{V}$ is diagonal.
The Schr\"{o}dinger equation \eqref{eq:SE} (with $\mathbb{W}=0$) then represents 16 uncoupled equations, and solving for all channels becomes a matter of individually finding the singlet and triplet partial wave phase shifts, $\delta^s_l$ and $\delta^t_l$ from the equations
\begin{equation}
  \left\{\frac{d^2}{dx^2} - \frac{l(l+1)}{x^2} + f(R)
  \left[ \mathcal{E} - \epsilon_{0,1}(x) \right] \right\} F_l^{s,t}(x) = 0 \ .
  \label{eq:SE-single}
\end{equation}

The asymptotic solution is
\begin{equation}
  \lim_{x\to\infty} F_l^{s,t}(x) = x e^{i\delta_l^{s,t}} \left[
    \cos\delta_l^{s,t} j_l(kx) - \sin\delta_l^{s,t} n_l(kx) \right] \ ,
\end{equation}
where $k=\sqrt{f(R)E}$.
We solve to some sufficiently large value $x_f$ and invert the matching condition to obtain
\begin{align}
  \tan\delta_l^{s,t} &= \frac{x_f j'_l(kx_f) - \beta^{s,t} j_l(kx_f)}
            {x_f n'_l(kx_f) - \beta^{s,t} n_l(kx_f)} \\
  \beta^{s,t} &= \frac{x_f (F_l^{s,t})'(x_f)}{F_l^{s,t}(x_f)} - 1 \ ,
\end{align}
where the primes denote derivatives with respect to $x$.
We use the same procedure (simplified to a single channel) from Appendix \ref{sec:full} to set the initial conditions and find the range $[x_i,x_f]$ in which the partial wave phase shift is convergent.

Once we have the singlet and triplet partial wave phase shifts, we transform back into the $\ket{F_A M_A F_B M_B}$ basis to find the cross sections for scattering between the channels listed in Table \ref{tab:channels}.
Many of the cross sections are redundant, so we group them according to the details of the scattering process.
There are five groups that represent $\Delta F=0$ scattering.
For $\sigma_0$, particles remain in the excited state and scatter from $M_{A,B}=\pm 1$ to $M_{A,B}=0$ (and vice versa).
For $\sigma_1$, particles remain in the ground or excited state with $M_{A,B}=0$.
For $\sigma_2$, one particle is in the ground state, and the other is in the excited state with any $M_{A,B}$.
For $\sigma_3$, particles remain in the excited state with $M_{A,B}=\pm 1$.
The exceptions are channels 15 and 16, which experience pure elastic scattering, each with a cross section $\sigma_4$.
The remaining cross sections have $\Delta F \neq 0$.
The cross section $\sigma_5$ represents $|\Delta F|=1$ scattering and is $p$-wave suppressed.
The cross section $\sigma_6$ represents $|\Delta F|=2$ scattering and is equal to $\sigma_0$.
The expressions for these cross sections are given as follows:
\begin{subequations}
\begin{align}
  \sigma_0 &= \frac{\pi}{2k^2} \sum_{l\textrm{ even}} (2l+1)
  \sin^2(\delta^s_l - \delta^t_l) \\
  \sigma_1 &= \frac{\pi}{2k^2} \sum_{l\textrm{ even}} (2l+1)
  \left[4\sin^2\delta^s_l + 12\sin^2\delta^t_l
    - 3\sin^2(\delta^s_l - \delta^t_l)\right] \\
  \sigma_2 &= \frac{\pi}{2k^2} \left[
    \sum_{l\textrm{ even}} (2l+1) 4\sin^2\delta^t_l
    + \sum_{l\textrm{ odd}} (2l+1)
    \left[2\sin^2\delta^s_l + 2\sin^2\delta^t_l
    - \sin^2(\delta^s_l - \delta^t_l)\right] \right] \\
  \sigma_3 &= \frac{\pi}{2k^2} \sum_{l} (2l+1)
  \left[2\sin^2\delta^s_l + 2\sin^2\delta^t_l
    - \sin^2(\delta^s_l - \delta^t_l)\right] \\
  \sigma_4 &= \frac{\pi}{2k^2} \sum_{l\textrm{ even}} (2l+1) 16\sin^2\delta^t_l \\
  \sigma_5 &= \frac{\pi}{2k^2} \sum_{l\textrm{ odd}} (2l+1)
  \sin^2(\delta^s_l - \delta^t_l) \\
  \sigma_6 &= \sigma_0 \ .
\end{align}
\end{subequations}
We form the total spin-averaged cross section by summing over all possible cross sections, weighted by a factor of $1/16$~\cite{Jamieson:2000yg}:
\begin{equation}
  \sigma_\textrm{tot} = \frac{\pi}{2k^2} \sum_l (2l+1)
  \begin{cases}
    \sin^2\delta^s_l + 9\sin^2\delta^t_l & l \textrm{ even} \\
    3\sin^2\delta^s_l + 3\sin^2\delta^t_l & l \textrm{ odd} \ .
  \end{cases}
  \label{eq:elastic-tot}
\end{equation}

\bibliography{physics-refs}

\end{document}